\documentclass[lettersize,journal]{IEEEtran}

\usepackage[caption=false,font=normalsize,labelfont=sf,textfont=sf]{subfig}
\usepackage{textcomp}
\usepackage{stfloats}
\usepackage{url}
\usepackage{verbatim}
\usepackage{graphicx}
\usepackage{cite}
\usepackage{titlesec}
\usepackage{enumitem}
\usepackage{hyperref}
\usepackage{amssymb}
\usepackage{algorithm}
\usepackage{algorithmic}

\usepackage{subcaption}

\usepackage{multirow}
\usepackage{makecell}

\usepackage{afterpage}
\usepackage{rotating}
\usepackage{subcaption}

\usepackage{lineno}

\usepackage{amsmath}
\usepackage{graphicx}
\usepackage{adjustbox}
\usepackage{xcolor}
\usepackage[utf8]{inputenc} % allow utf-8 input
\usepackage[T1]{fontenc}    % use 8-bit T1 fonts
\usepackage{hyperref}       % hyperlinks
\usepackage{url}            % simple URL typesetting
\usepackage{booktabs}       % professional-quality tables
\usepackage{amsfonts}       % blackboard math symbols
\usepackage{nicefrac}       % compact symbols for 1/2, etc.
\usepackage{microtype}      % microtypography
\usepackage{lipsum}
\usepackage{graphicx}
\graphicspath{{media/}}     % organize your images and other figures under media/ folder
\usepackage{float}
\usepackage{multirow}
\usepackage{orcidlink}
\begin{document}

\title{Automated Hardware Validation Test Plan Generation for Large Scale AI Datacenter Platforms Using a Generative AI Multi-Agents Architecture}

\author{Mohammed-Khalil Ghali~\orcidlink{0009-0003-5109-7084},
        Saurabh Kulkarni\textsuperscript{*}~\orcidlink{0009-0009-6720-1126},
        Prathamesh Kulkarni~\orcidlink{0009-0006-1499-8766},
        Rohan Kulkarni~\orcidlink{0009-0007-1398-4327},
        Sangwon Yoon~\orcidlink{0000-0002-1613-0745},
        Daehan Won~\orcidlink{0000-0002-2566-8061}
        %%Corresponding author TBD\textsuperscript{*}~\orcidlink{0000-0002-2566-806x}%
\thanks{* Corresponding author: Saurabh Kulkarni (saurabhkulkarni@meta.com)}%
\thanks{Mohammed-Khalil Ghali, Sangwon Yoon and Daehan Won (Emails: mghali1@binghamton.edu, yoons@binghamton.edu, dhwon@binghamton.edu) are with the School of Systems Science and Industrial Engineering, Binghamton University, State University of New York, Binghamton, NY 13902, USA.}
\thanks{Saurabh Kulkarni, Prathamesh Kulkarni and Rohan Kulkarni (Emails: saurabhkulkarni@meta.com, prathkulkarni@meta.com, irohan@meta.com) are with Meta Platforms, Inc. 1 Hacker Way, Menlo Park, CA 94025, USA.}

}
\markboth{arXiv}%
{Shell \MakeLowercase{\textit{et al.}}: A Sample Article Using IEEEtran.cls for IEEE Journals}

\IEEEpubid{}

\maketitle

\begin{abstract}
Large-scale AI datacenter platforms comprise thousands of heterogeneous hardware components whose validation requires comprehensive fault injection test plans. Today, these plans are authored manually: engineers review hardware self-healing validation documents and bills of materials, enumerate failure modes per field-replaceable unit, and produce flat lists of single-layer test cases. This process is labor-intensive, error-prone, and heavily dependent on institutional knowledge. Coverage gaps are discovered late, traceability between test cases and source specifications is implicit at best, and the effort must be substantially repeated for each new platform. This paper presents a generative AI multi-agent architecture that automates the generation of structured hardware validation test plans from two canonical engineering inputs: hardware self-healing validation documents, which enumerate known failure modes and their associated detection and remediation behaviors for each field-replaceable unit, and components Bills of Material (BOM). An ingestion agent normalizes heterogeneous input formats into a canonical intermediate representation. A classification agent maps component entries to functional domains using contextual reasoning over part descriptions and sub-category hierarchies. A generation agent synthesizes test cases by combining normalized failure modes with domain-classified component data, filling specification gaps and producing edge-case scenarios. The output conforms to a standardized schema suitable for direct import into an internal automated validation software. The framework is evaluated on two production platforms, comparing auto-generated test plans against manually authored baselines. Results demonstrate significant coverage expansions of 74.2\% and 51.4\% over manual baselines, effectively reducing the authoring burden from days of manual review to hours of draft refinement. The system yields fully traceable mappings from each test case to its originating specification entries, and the multi-agent decomposition provides a reusable abstraction portable across hardware platform generations. Automated panel and human expert evaluations confirm 100\% extraction fidelity for existing specifications and high acceptance rates for newly generated scenarios, validating the framework as a robust human-in-the-loop force multiplier.
\end{abstract}


\section{Introduction}
\IEEEPARstart{T}{he}
\label{intro}
global demand for artificial intelligence capability is driving an unprecedented expansion of datacenter infrastructure. Hyperscale operators are deploying AI-optimized platforms at rack and cluster scale, integrating specialized accelerators, high-bandwidth interconnects, liquid cooling systems, and power delivery architectures that differ fundamentally from prior-generation compute infrastructure. Deployment timelines have compressed sharply: platforms that once followed multi-year qualification cycles are now expected to move from design validation to production readiness in months. The competitive pressure to bring training and inference capacity online has made speed of deployment a first-order business constraint, yet the hardware beneath these platforms must still meet stringent reliability and availability targets. A single undetected fault class in a rack-scale system can cascade across hundreds of coordinated nodes, disrupting large-scale training and inference jobs incurring significant cost in both compute time and operational recovery.

Hardware self-healing validation which is the systematic process of verifying that a platform behaves and self heals correctly under fault conditions is the primary gate between engineering release and production deployment. For AI datacenter hardware, validation centers on fault injection testing: deliberately introducing component-level and system-level faults to verify detection, isolation, and remediation behavior. The scope of this testing is determined by a test plan, a structured document that specifies which faults to inject, on which components, under what conditions, and with what expected outcomes. The quality and completeness of the test plan directly governs the residual risk carried into production.

Despite its criticality, test plan authoring remains a predominantly manual process. Engineers consult hardware self-healing validation documents—structured specifications that enumerate known failure modes, detection mechanisms, and remediation behaviors for each field-replaceable unit (FRU), serving as the primary reference for verifying that the platform's self-healing logic correctly detects, isolates, and recovers from hardware faults across the deployed fleet—and cross-reference them against the platform's bill of materials (BOM) to identify which components require coverage. Test cases are then written individually, typically in spreadsheet form, and reviewed through engineering judgment. This workflow suffers from several structural deficiencies that scale poorly with platform complexity. Each test case is hand-authored from self-healing validation documents that vary in format, schema, and completeness across component vendors and internal teams; for platforms comprising hundreds of distinct FRUs, initial test plan generation alone can consume weeks of engineering effort, creating a direct bottleneck on validation schedules that are already compressed by accelerated deployment timelines. The mapping between a test case and the specification entry that motivated it is rarely recorded explicitly, so when test plans are reviewed, updated, or ported to a new platform the rationale for inclusion or exclusion of specific scenarios is lost, making coverage auditing subjective and gap identification unreliable. Because test plans encode platform-specific component identifiers and tribal knowledge of prior failure behavior, they cannot be transferred meaningfully to new platforms, and each generation requires near-complete re-authoring even when the underlying failure mode taxonomy and functional architecture share substantial overlap with predecessors. Without a systematic method to identify under-tested components or missing fault categories, coverage gaps are typically discovered during test execution or, worse, after deployment through field incidents, and the feedback loop from gap discovery to test plan remediation remains slow and reactive.

The emergence of agentic AI architectures, which are systems composed of multiple specialized language model agents that collaborate through structured interfaces to accomplish complex reasoning tasks, offers a new paradigm for automating knowledge-intensive engineering workflows. Unlike monolithic model approaches, multi-agent systems decompose a problem into discrete reasoning stages, each handled by an agent optimized for a specific subtask. This decomposition mirrors the natural structure of engineering processes: distinct phases of data ingestion, normalization, classification, synthesis, and validation, each requiring different domain context and reasoning strategies. Recent advances in large language models have demonstrated strong performance on structured document understanding, schema inference, and domain-aware generation tasks, making agent-based pipelines a viable approach for engineering automation problems that were previously intractable without deep domain-specific software.

This paper presents a multi-agent generative AI system that automates hardware validation test plan generation from self-healing validation documents and component bills of materials. The system replaces manual authoring with a pipeline of specialized agents: a classification agent that maps BOM components to functional domains through contextual reasoning, an ingestion agent that normalizes heterogeneous specification formats into a canonical schema, and a generation agent that synthesizes test cases by combining failure mode data with domain classified component inventories. The output is a structured, traceable test plan conforming to a standardized schema suitable for import into an internal validation management platform. The evaluation strategy of the framework is on production rack-scale AI infrastructure and compare auto-generated plans against manually authored baselines across coverage, traceability, and authoring efficiency metrics.

\begin{figure*}[!t]
    \centering
    \includegraphics[width=0.8\textwidth]{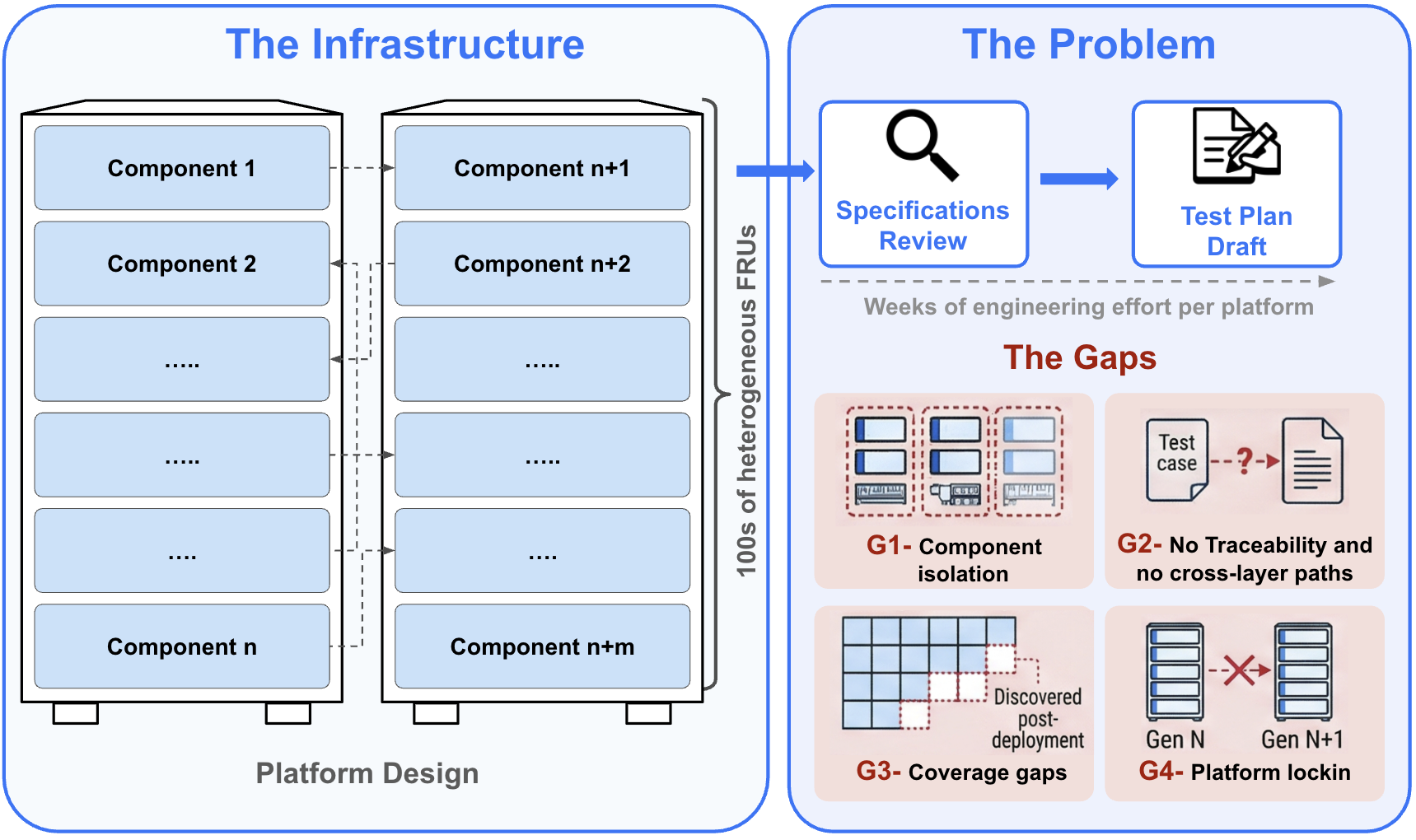}
    \caption{Summary of key challenges in manual hardware validation test plan authoring for AI datacenter platforms}\label{fig:problems}
\end{figure*}

  Fig. \ref{fig:problems} summarizes the key challenges associated with manual test plan authoring.

The remainder of this paper is organized as follows. Section \ref{relatedwork} surveys related work in hardware validation automation, fault injection methodology, and Large Language Model-based engineering tools. Section \ref{methdology} details the proposed methodology. Section \ref{sec:experimental_setup} summarizes the experimental setup. Section \ref{sec:results} presents the proposed framework's results. Section \ref{conclusion} concludes, while providing limitations and future directions.

\section{Related Work}
\label{relatedwork}

\begin{table*}[t]
  \centering
  \caption{Summary of related work, main gaps, and proposed contributions}
  \label{tab:related_work}
  \renewcommand{\arraystretch}{1.4}
  \begin{tabular}{>{\raggedright\arraybackslash}p{3.5cm} p{5.5cm} p{5.5cm}}
  \hline
  \textbf{Current Methods} & \textbf{Main Gaps} & \textbf{Proposed Contributions} \\
  \hline
  Fault Injection \cite{pourreza2023survey, yu2026survey, cinque2025cosmos, opara2025chaos} & Scenarios are manually defined; no link to design specifications. No automated synthesis of test plans from engineering artifacts. & Specification-driven test case generation
  from BOM and self-healing validation document inputs. Full traceability from each engineering input to each test case. \\
  \hline
  Runtime Reliability Automation \& Digital Twins \cite{Ray2026TRiSM, Zhang2026Infrastructure, morgan2025digital} & Operate post-deployment; design-time validation not
  addressed. Consume system telemetry, not structured engineering artifacts. Produce behavioral simulations, not discrete test plans. & Design-time validation planning prior to
  hardware deployment. Structured artifacts (BOM, self-healing validation document) as direct system inputs. Outputs structured, executable hardware validation test plans. \\
  \hline
  Engineering Automation \cite{Hao2026Survey, Uslu2026Orchestration} & Monolithic models with limited domain-specific grounding. No traceability between outputs and source
  engineering inputs. Not designed for structured test plan generation. & Domain-aware prompting grounded in hardware taxonomy and failure mode knowledge. Schema-driven outputs
  with explicit citation of source artifacts. Purpose-built pipeline for hardware validation test plan synthesis. \\
  \hline
  Generative AI Agents \& Multi-Agent Architectures \cite{Ivanov2026Agentic, Liu2026STAR, yu2025systematic} & Focused on runtime orchestration; no structured engineering
  workflow. Lack domain constructs such as failure mode taxonomies or component hierarchies. No consistency guarantees or coverage assurance on outputs. & Multi-agent pipeline
  decomposed into ingestion, classification, and test generation stages. Cross-agent consistency checks and structured inter-agent communication. Coverage verification against
  full input component and failure mode scope. \\
  \hline
  \textbf{The proposed model} & Real-time multi-agent pipeline that fuses BOM ingestion, self-healing validation document failure extraction, domain classification, and automated test case
  generation. & Enables specification-driven hardware validation test plan synthesis with full traceability and human-in-the-loop quality assurance.
  \\
  \hline
  \end{tabular}
  \end{table*}
\subsection{Overview}

Hardware validation in large-scale AI datacenter systems draws on three research areas: fault injection and reliability testing, engineering automation, and Large Language Model-based multi-agent systems. While each area has matured independently, their combined application to systematic, artifact-driven validation planning remains largely unexplored. Modern AI clusters present unique challenges, such as interdependent GPU memory fabrics, high-bandwidth interconnects like NVLink or InfiniBand, and specialized liquid-cooling units, which often exceed the capabilities of general-purpose cloud validation. Most prior work addresses runtime reliability or low-level software fault modeling, leaving a clear gap in design-time test planning driven by structured engineering inputs such as Bills of Materials (BOMs) and self-healing validation documents.

%% ---------------------------------------------------------------
\subsection{Fault Injection and Reliability Validation}
%% ---------------------------------------------------------------

Fault injection is a well-established technique for assessing system dependability in distributed and cloud environments. Traditional surveys in IEEE and ACM venues document injection techniques spanning hardware, virtualization, and application layers, reflecting the growing complexity of modern datacenter systems~\cite{pourreza2023survey, yu2026survey}. However, as clusters transition from general-purpose CPU compute to highly accelerated AI nodes, the blast radius of hardware faults has shifted, necessitating more granular isolation testing to prevent cluster-wide outages.

Recent work has extended fault injection to cloud-native and AI-driven systems to address these shifts. Cinque et al.~\cite{cinque2025cosmos} present COSMOS, a hardware-assisted fault injection framework that uses nested virtualization to test hardware-assisted hypervisors such as KVM, Xen, and Jailhouse. By injecting faults without modifying the target system, COSMOS exposes non-fail-stop behaviors, including silent data corruption or gray failures, which are particularly catastrophic for large-scale distributed training runs. Related studies on fault injection in microservice and edge environments further highlight the challenge of validating resilience across heterogeneous deployments where hardware-software dependencies are tightly coupled~\cite{yu2025systematic}.

Chaos engineering has emerged as a complementary operational approach, enabling continuous failure testing in production systems. Recent work describes AI-enhanced and policy-driven chaos frameworks that orchestrate complex multi-cloud failure scenarios at scale~\cite{opara2025chaos}. Despite these advances, both fault injection and chaos engineering remain primarily experiment-driven rather than plan-driven. In these paradigms, test scenarios are often defined manually based on engineering intuition, coverage is typically limited to individual software components, and there is no systematic mechanism to derive specific hardware test cases directly from design-time engineering specifications.

%% ---------------------------------------------------------------
\subsection{Automation in Reliability and Systems Engineering}
%% ---------------------------------------------------------------

Machine learning and generative AI have been applied to system observability, root cause analysis, and operational automation~\cite{Ray2026TRiSM}. These approaches, often categorized under AIOps, improve runtime responsiveness by identifying patterns in log files and telemetry streams. However, these models operate downstream of the hardware design, meaning they react to failures rather than proactively structuring the validation process based on the physical architecture of the system.

Digital twin frameworks extend this automation by enabling simulation-based reliability assessment. Recent literature discusses digital twin models that simulate cascading failures across critical infrastructure and cyber-physical systems~\cite{Zhang2026Infrastructure, morgan2025digital}. While digital twins provide sophisticated behavioral simulations, their utility in hardware validation is often bottlenecked by the fidelity-complexity trade-off. Their outputs are behavioral predictions rather than the discrete validation artifacts, such as structured hardware test plans or executable scripts, required by test engineers during the hardware bring-up and qualification phase.

Large Language Model-based systems have also been explored for engineering workflow automation, including configuration management and network orchestration~\cite{Hao2026Survey, Uslu2026Orchestration}. However, these applications remain focused on operational tasks and do not address the structural reasoning required to map a multi-level hardware hierarchy to a comprehensive test suite. A consistent limitation across this body of work is the absence of integration between structured engineering inputs and automated reasoning systems. Automation remains disconnected from the formal design phase, leading to a persistent silo between hardware design teams and validation engineers.

%% ---------------------------------------------------------------
\subsection{Multi-Agent Systems}
%% ---------------------------------------------------------------

Multi-agent systems based on Large Language Models represent the frontier of complex reasoning and task decomposition. Recent work proposes agentic architectures in which multiple specialized agents collaborate on tasks beyond the capacity of monolithic models, offering improved modularity and interpretability in industrial and cyber-physical domains~\cite{Ivanov2026Agentic, Liu2026STAR}. In cloud and network management, distributed agent frameworks have been applied to fault diagnosis, traffic orchestration, and decision support~\cite{Uslu2026Orchestration}. By leveraging specialized personas, these systems can simulate organizational workflows, yet their application to the rigorous requirements of hardware engineering remains in its infancy.

Despite this progress, current multi-agent systems suffer from several primary deficiencies when applied to the context of hardware validation. First, most multi-agent frameworks rely on general-purpose reasoning and do not natively incorporate hardware-centric constructs, such as Failure Mode Taxonomies or the physical hierarchies inherent in complex AI cluster architectures. Second, current research prioritizes finding a singular solution or response rather than ensuring the exhaustive traceability and coverage guarantees required for engineering certification. There is currently no formal mechanism to ensure an agentic workflow has addressed every critical component listed in a design specification. Finally, these systems are not designed to generate the heterogeneous, structured artifacts required for validation, such as formal test procedures, risk assessments, and compliance documentation. These limitations restrict the applicability of multi-agent systems to structured engineering workflows, highlighting a pressing need for frameworks that can ingest structured hardware data and, through specialized collaboration, produce a rigorous and traceable validation plan.

 Table~\ref{tab:related_work} summarizes the key previous work from the reviewed literature, the main limitations of each, and how the proposed framework addresses them.
%% ---------------------------------------------------------------
%% ---------------------------------------------------------------

The reviewed literature reveals four recurring limitations that motivate this work:

\begin{itemize}
    \item Fault injection rely on manually defined scenarios and cannot automatically generate test cases from engineering specifications.
    \item Automation techniques operate at runtime and do not support design-time validation planning.
    \item Large Language Models (LLMs) and multi-agent systems have not been applied to structured engineering artifact generation with explicit traceability to source inputs.
\end{itemize}

This paper addresses these gaps by proposing a multi-agent generative AI framework that transforms structured engineering inputs specifically BOMs and self-healing validation documents into comprehensive, traceable hardware self-healing validation test plans. Unlike prior work, the proposed approach integrates domain-aware reasoning and schema-driven output generation within a unified multi-agent architecture.

%% ---------------------------------------------------------------
%% ---------------------------------------------------------------

\section{Methodology}
\label{methdology}

\begin{figure*}[t]
    \centering
    \includegraphics[width=0.8\textwidth]{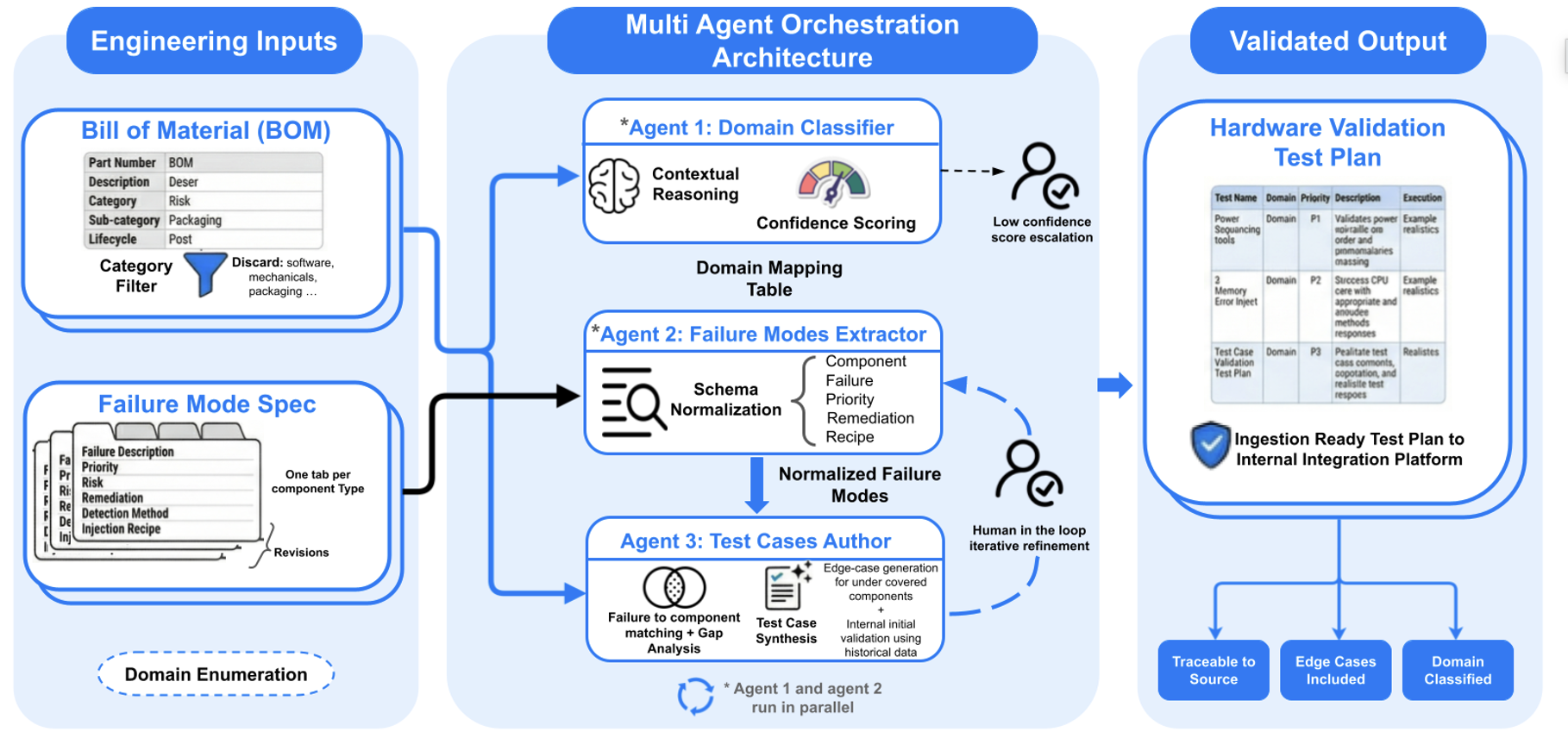}
    \caption{Multi-agent generative AI framework for automated hardware self-healing validation test plan generation showing domain-aware component classification, failure mode extraction, and test case synthesis}
        \label{graphicalabstract}
\end{figure*}

Fig. \ref{graphicalabstract} describes the proposed multi-agent generative AI framework for automated hardware validation test plan generation from two canonical engineering inputs: self-healing validation documents and BOM. The architecture decomposes the generation task into three specialized agents: a domain classifier agent that maps BOM entries to functional domains using contextual reasoning, a failure modes extractor agent that normalizes heterogeneous failure mode specifications into a canonical intermediate representation, and a test cases author agent that synthesizes test cases by combining normalized failure modes with domain-classified component data. The domain classifier agent and the failure modes extractor agent operate in parallel as they process independent inputs with no data dependencies between them; their outputs are subsequently consumed by the test cases author agent. Each of these components are described in details in the following subsections.

\subsection{Domain Classifier Agent}

The first stage of the framework addresses the classification of bill of materials entries into functional hardware domains. In rack-scale AI platforms, a bill of materials typically contains hundreds of distinct component entries spanning diverse functional areas including compute accelerators, power delivery, thermal management, storage, networking, and system management. Grouping these components into coherent functional domains is essential for organizing test plans by engineering discipline, identifying coverage gaps at the domain level, and assigning domain-appropriate test case attributes in the final output. Rather than relying on static rule-based mappings that require manual maintenance as component taxonomies evolve, the proposed approach employs an LLM to dynamically propose a domain taxonomy and classify each component based on contextual reasoning over its description and sub-category designation.

Let $\mathbf{B} = \{b_1, b_2, \ldots, b_N\}$ denote the set of bill of materials entries, where $N$ is the total number of components. Each entry $b_i$ is characterized by a tuple:
\begin{equation}
b_i = (d_i, s_i, \text{cat}_i, \delta_i)
\end{equation}
where $d_i$ is the textual description of the component, $s_i$ is the sub-category label, $\text{cat}_i$ is the top-level category, and $\delta_i$ encapsulates additional attributes such as field-replaceability status and quantity per rack.

Certain top-level categories are not relevant to hardware fault injection validation, such as software licenses, mechanical assemblies, or packaging materials. A configurable discard set $\mathcal{C}_{\text{disc}}$ defines the categories to exclude prior to classification. The filtered bill of materials is computed as:
\begin{equation}
\mathbf{B}' = \{ b_i \in \mathbf{B} \mid \text{cat}_i \notin \mathcal{C}_{\text{disc}} \}
\end{equation}
where $\mathbf{B}'$ contains only the hardware components relevant to validation test planning.

To provide the language model with sufficient context for classification, the unique sub-categories are extracted from the filtered bill of materials along with representative sample descriptions. Let $\mathcal{U} = \{u_1, u_2, \ldots, u_M\}$ denote the set of unique sub-category values, formally defined as:
\begin{equation}
\mathcal{U} = \{ s_i \mid b_i \in \mathbf{B}' \}
\end{equation}
For each unique sub-category $u \in \mathcal{U}$, a bounded set of sample descriptions is collected to provide the classifier with concrete examples of parts within that grouping:
\begin{equation}
\mathcal{X}(u) = \{ d_i \mid b_i \in \mathbf{B}', s_i = u \}_{1:P}
\end{equation}
where the subscript $1{:}P$ indicates that at most $P$ sample descriptions are retained per sub-category to bound the input context size.

To guide the classification, a platform-specific enumeration list of candidate domains $\mathcal{D}_{\text{enum}} = \{D_1, D_2, \ldots, D_Q\}$ is defined for each hardware platform under validation. This enumeration reflects the functional architecture and engineering disciplines relevant to the target platform and is provided as input to the language model to constrain the classification space. The classification is performed by a language model that receives the full set of sub-categories with their sample descriptions alongside the platform-specific domain enumeration, and produces three outputs simultaneously: the final domain taxonomy, a classification mapping, and a confidence assessment. This is formally expressed as:
\begin{equation}
(\mathcal{D}, \mu_{\text{dom}}, \kappa) = f_{\text{LLM}}^{\text{class}}(\mathcal{U}, \mathcal{X}, \mathcal{D}_{\text{enum}})
\end{equation}
where $\mathcal{D} \subseteq \mathcal{D}_{\text{enum}}$ is the set of domains that received at least one assignment, $\mu_{\text{dom}}: \mathcal{U} \rightarrow \mathcal{D}_{\text{enum}}$ is the classification mapping that assigns each sub-category to exactly one domain from the enumeration, and $\kappa: \mathcal{U} \rightarrow [0, 100]$ is the confidence function that quantifies the certainty of each classification decision.

The confidence scoring follows a three-tier interpretation:
\begin{equation}
\text{confidence level} =
\begin{cases}
\text{high} & \text{if } \kappa(u) \in [70, 100] \\
\text{moderate} & \text{if } \kappa(u) \in [50, 70) \\
\text{low} & \text{if } \kappa(u) < 50
\end{cases}
\end{equation}
where high-confidence assignments indicate clear functional alignment between the sub-category and the proposed domain, moderate-confidence assignments flag entries that warrant human review, and low-confidence assignments indicate that the sub-category description was too ambiguous for reliable classification.

The domain assignment is propagated from the sub-category level to each individual bill of materials entry. For each component $b_i \in \mathbf{B}'$, the assigned domain is given by:
\begin{equation}
\text{dom}(b_i) = \mu_{\text{dom}}(s_i)
\end{equation}
and the associated confidence is:
\begin{equation}
\text{conf}(b_i) = \kappa(s_i)
\end{equation}

The output of this stage is a domain-enriched bill of materials $\mathbf{B}^*$ where each entry is augmented with its functional domain label and classification confidence, formally defined as:
\begin{equation}
\mathbf{B}^* = \left\{ (b_i, \text{dom}(b_i), \text{conf}(b_i)) \mid b_i \in \mathbf{B}' \right\}
\end{equation}

This domain-enriched representation serves two purposes in downstream processing: it provides the functional domain label required by the output test case schema, and it enables the test cases author agent to reason about component relationships within and across domains during cross-component test synthesis.

\subsection{Failure Modes Extractor Agent}

The second stage of the proposed framework, which operates in parallel with the domain classifier agent, addresses the challenge of extracting structured failure mode data from heterogeneous self-healing validation documents. In production environments, self-healing validation documents are organized in several formats depending on the team working on it among which organizing as multi-tab spreadsheet documents where each tab corresponds to a distinct hardware component or subsystem. These tabs vary in column naming conventions, header row placement, and schema structure across component vendors and internal engineering teams, making automated extraction non-trivial. The failure modes extractor agent normalizes these heterogeneous inputs into a canonical intermediate representation suitable for downstream processing by the test cases author agent.

Let $\mathcal{T} = \{T_1, T_2, \ldots, T_L\}$ denote the complete set of tabs within a self-healing validation document, where $L$ is the total number of tabs. Not all tabs contain actionable failure mode data; some contain metadata, revision history, or unrelated reference material. A domain-relevant whitelist $\mathcal{W} \subset \mathcal{T}$ defines the subset of tabs that contain hardware failure mode information. The filtered set of valid tabs is formally expressed as:
\begin{equation}
\mathcal{T}_{\text{valid}} = \{ T_j \in \mathcal{T} \mid T_j \in \mathcal{W} \}
\end{equation}
where each $T_j \in \mathcal{T}_{\text{valid}}$ corresponds to a specific hardware component or subsystem containing failure mode entries that require extraction.

For each valid tab $T_j$, the raw content is represented as a tabular matrix $\mathbf{R}_j \in \mathbb{R}^{n_j \times m_j}$, where $n_j$ denotes the number of rows and $m_j$ denotes the number of columns. Since header row placement is inconsistent across tabs, a header detection function $g_{\text{header}}$ is applied to identify the row containing column descriptors:
\begin{equation}
\mathbf{h}_j = g_{\text{header}}(\mathbf{R}_j)
\end{equation}
where $\mathbf{h}_j = [h_{j,1}, h_{j,2}, \ldots, h_{j,m_j}]$ is the detected header vector. If the initially detected header row contains no meaningful non-empty cells, the function automatically falls back to the subsequent row, which ensures robustness against blank or formatting-only rows at the top of each tab.

The component name associated with each tab is inferred from its title using an LLM. Rather than relying on brittle string-matching heuristics, the language model interprets the tab title in the context of hardware engineering terminology. For each valid tab $T_j$, the component name is computed as:
\begin{equation}
c_j = f_{\text{LLM}}^{\text{comp}}(\text{title}(T_j))
\end{equation}
where $f_{\text{LLM}}^{\text{comp}}$ denotes the language model function specialized for component name inference, and $\text{title}(T_j)$ returns the human-readable title of tab $T_j$.

The central challenge of ingestion lies in mapping the heterogeneous column headers of each tab to a unified canonical schema. Let $\mathcal{F} = \{f_1, f_2, \ldots, f_K\}$ denote the set of canonical schema fields that define the normalized failure mode representation. These fields include failure description, failure category, priority, risk assessment, remediation procedure, detection method, and error injection recipe. The column mapping for each tab is performed by a language model that receives the detected headers and the target schema, formally expressed as:
\begin{equation}
\mu_j: \mathbf{h}_j \rightarrow \mathcal{F}
\end{equation}
\begin{equation}
\mu_j = f_{\text{LLM}}^{\text{map}}(\mathbf{h}_j, \mathcal{F})
\end{equation}
where $\mu_j$ is a partial injective mapping from detected column headers to canonical schema fields. The mapping is partial because not every column in the source tab corresponds to a schema field, and injective because each schema field maps to at most one source column.

Using the inferred component name $c_j$ and the column mapping $\mu_j$, each data row $\mathbf{r}_{j,i}$ (for $i = 1, 2, \ldots, n_j'$, where $n_j'$ is the number of data rows excluding the header) is normalized into a canonical failure mode record:
\begin{equation}
\phi_{j,i} = \mathcal{N}(\mathbf{r}_{j,i}, \mu_j, c_j)
\end{equation}
where $\mathcal{N}$ is the normalization function that extracts the value at each mapped column index and assigns it to the corresponding schema field, while setting the component field to $c_j$. Fields for which no source column exists in tab $T_j$ are preserved as empty values to avoid introducing fabricated information. This design choice is critical: the failure modes extractor agent performs no gap-filling or content generation, which ensures that the normalized output faithfully represents the source specification.

The complete set of normalized failure modes aggregated across all valid tabs is given by:
\begin{equation}
\Phi = \bigcup_{T_j \in \mathcal{T}_{\text{valid}}} \left\{ \phi_{j,i} \right\}_{i=1}^{n_j'}
\end{equation}
where $\Phi$ constitutes the canonical failure mode repository that serves as a primary input to the test cases author agent.

\subsection{Test Cases Author Agent}

The third and central stage of the framework synthesizes structured hardware validation test cases by combining the normalized failure modes from the failure modes extractor agent with the domain-classified component inventory from the domain classifier agent. The test cases author agent performs four distinct functions: it maps bill of materials components to their corresponding self-healing validation document coverage, identifies and fills gaps in existing failure mode records, generates new test cases for under-covered or uncovered components, and produces cross-component interaction test scenarios that exercise failure propagation paths across functionally coupled subsystems. The design follows a detection-path-first principle, where test cases are derived by enumerating the monitoring and detection pathways available for each component and validating that each pathway functions correctly under fault conditions.

\subsubsection{Coverage Mapping}

The first operation establishes a mapping between the physical component inventory and the failure mode specifications to determine which components have existing test coverage and which represent coverage gaps. Let $\Phi$ denote the normalized failure mode set from the failure modes extractor agent and $\mathbf{B}^*$ the domain-enriched bill of materials from the domain classifier agent. The coverage mapping function is defined as:
\begin{equation}
\mathcal{C}: \mathbf{B}^* \rightarrow 2^{\Phi} \cup \{\emptyset\}
\end{equation}
where $\mathcal{C}(b_i)$ returns the subset of failure modes in $\Phi$ whose component field semantically corresponds to bill of materials entry $b_i$, or the empty set if no self-healing validation document coverage exists for that component. This mapping employs fuzzy semantic matching to account for naming variations between bill of materials descriptions and self-healing validation document component identifiers.

Based on this mapping, the bill of materials entries are partitioned into two disjoint sets:
\begin{align}
\mathbf{B}^*_{\text{covered}} &= \{ b_i \in \mathbf{B}^* \mid \mathcal{C}(b_i) \neq \emptyset \} \\
\mathbf{B}^*_{\text{zero}} &= \{ b_i \in \mathbf{B}^* \mid \mathcal{C}(b_i) = \emptyset \}
\end{align}
where $\mathbf{B}^*_{\text{covered}}$ contains components with at least one existing failure mode in the self-healing validation document, and $\mathbf{B}^*_{\text{zero}}$ contains components with no existing coverage that require new test case generation.

To avoid redundant test generation for functionally identical components that differ only in minor attributes such as cable length or form factor variant, similar bill of materials entries in $\mathbf{B}^*_{\text{zero}}$ are grouped into logical component groups:
\begin{equation}
\mathcal{G}_{\text{zero}} = \text{group}(\mathbf{B}^*_{\text{zero}})
\end{equation}
where each group $g \in \mathcal{G}_{\text{zero}}$ represents a set of bill of materials entries that share the same functional role and should be covered by a single set of test cases.

\subsubsection{Gap Analysis and Completion}

For each existing failure mode $\phi \in \Phi$, the gap analysis identifies schema fields that are present in the canonical representation but contain empty values due to missing data in the source specification. The set of missing fields for a given failure mode is defined as:
\begin{equation}
\mathcal{G}(\phi) = \{ f_k \in \mathcal{F} \mid \phi[f_k] = \emptyset \}
\end{equation}
where $\phi[f_k]$ denotes the value of field $f_k$ in failure mode record $\phi$.

For each failure mode where $\mathcal{G}(\phi) \neq \emptyset$, a language model fills the missing fields based on the component type, failure description, and any available non-empty fields. The gap-filled failure mode is computed as:
\begin{equation}
\hat{\phi} = f_{\text{LLM}}^{\text{fill}}(\phi, \mathcal{G}(\phi))
\end{equation}
where $f_{\text{LLM}}^{\text{fill}}$ generates values only for the fields in $\mathcal{G}(\phi)$ while preserving all existing non-empty values. This constraint ensures that original specification data is never overwritten by generated content. The updated failure mode set after gap completion is:
\begin{equation}
\hat{\Phi} = \left\{ \hat{\phi}_k \mid \phi_k \in \Phi \right\}
\end{equation}

\subsubsection{Edge Case and Coverage Expansion}

For components with thin coverage (few existing failure modes) and for zero-coverage component groups, the test cases author agent synthesizes new test cases using a detection-path-first reasoning strategy. Rather than enumerating hypothetical physical failures, the agent identifies the monitoring, telemetry, and management pathways available for each component and generates test cases that validate the correct functioning of each detection path under injected fault conditions.

For each component or zero-coverage group requiring new test cases, the language model receives the component identity, relevant bill of materials context, and any existing failure modes to avoid duplication. The generation is formally expressed as:
\begin{equation}
\Phi_{\text{new}}(b_i) = f_{\text{LLM}}^{\text{gen}}(b_i, \mathcal{C}(b_i), \mathcal{X}_{\text{BOM}}(b_i))
\end{equation}
where $\mathcal{X}_{\text{BOM}}(b_i)$ provides the bill of materials descriptions for contextual grounding, and each generated failure mode $\phi_{\text{new}} \in \Phi_{\text{new}}(b_i)$ must satisfy the tooling-testability constraint:
\begin{equation}
\forall \phi_{\text{new}} \in \Phi_{\text{new}}(b_i): \quad \phi_{\text{new}}[\text{injection}] \neq \emptyset \;\wedge\; \phi_{\text{new}}[\text{detection}] \neq \emptyset
\end{equation}
This constraint enforces that every generated test case specifies both a concrete software-injectable fault condition and a verifiable detection pathway through the monitoring infrastructure, thereby excluding test scenarios that would require physical hardware manipulation.

\subsubsection{Cross-Component Interaction Test Synthesis}

In addition to single-component test cases, the framework generates cross-component interaction tests that validate the end-to-end detection and response chain when a fault in one component triggers cascading effects in functionally coupled subsystems.

Let $\Phi^* = \hat{\Phi} \cup \bigcup_{b_i} \Phi_{\text{new}}(b_i)$ denote the complete set of single-component failure modes after gap filling and coverage expansion. The cross-component test generation function takes the complete failure mode set and the bill of materials context as input:
\begin{equation}
\mathcal{I} = f_{\text{LLM}}^{\text{cross}}(\Phi^*, \mathbf{B}^*)
\end{equation}
where $\mathcal{I} = \{\iota_1, \iota_2, \ldots, \iota_R\}$ is the set of cross-component interaction test cases, and each test $\iota_r$ specifies a trigger component, an injection method, the expected downstream response in one or more affected components, and the validation procedure to confirm the response chain. Each cross-component test case encodes a directed interaction:
\begin{equation}
\iota_r = (b_{\text{src}}, b_{\text{dst}}, \phi_{\text{trigger}}, \phi_{\text{response}}, v_r)
\end{equation}
where $b_{\text{src}}$ is the source component where the fault is injected, $b_{\text{dst}}$ is the downstream component where the cascading effect is observed, $\phi_{\text{trigger}}$ describes the injected fault condition, $\phi_{\text{response}}$ describes the expected detection and remediation response, and $v_r$ specifies the validation procedure. All cross-component tests are assigned the highest priority classification given that undetected cascading failures represent the highest-risk scenario in production deployments.

\subsubsection{Test Plan Composition and Output Formatting}

The final test plan is composed by aggregating all test case categories into a unified, structured output. The complete test plan $\mathcal{P}$ is defined as:
\begin{equation}
\mathcal{P} = \hat{\Phi} \cup \Phi_{\text{new}} \cup \mathcal{I}
\end{equation}
where $\hat{\Phi}$ contains the original failure modes with gaps filled, $\Phi_{\text{new}} = \bigcup_{b_i} \Phi_{\text{new}}(b_i)$ contains all newly generated single-component test cases, and $\mathcal{I}$ contains the cross-component interaction tests.

Each test case $p \in \mathcal{P}$ is transformed into the output schema required by the downstream internal automated validation platform. The domain assignment for each test case is determined by its corresponding bill of materials mapping:
\begin{equation}
\text{domain}(p) =
\begin{cases}
\text{dom}(b_i) & \text{if } p \in \hat{\Phi} \cup \Phi_{\text{new}} \text{ and } \exists\, b_i \in \mathbf{B}^* \\
D_{\text{cross}} & \text{if } p \in \mathcal{I}
\end{cases}
\end{equation}
where $\text{dom}(b_i)$ is the domain assigned by the domain classifier agent, and $D_{\text{cross}}$ is a designated domain label for cross-component interaction tests. The final structured output for each test case conforms to the following schema transformation:
\begin{equation}
\text{output}(p) = \mathcal{O}(p, \text{domain}(p))
\end{equation}
where $\mathcal{O}$ maps each test case and its domain to a structured record containing a descriptive name, domain label, priority level, composite test case description, and execution system designation. Each generated test case is tagged with its provenance—whether originating from the source specification or generated by the test cases author agent—to maintain full traceability between the test plan and its originating inputs.

\subsection{Implementation and Reproducibility}

  The proposed multi-agent framework was developed and executed using agentic AI orchestration tools built internally at Meta. The underlying language models were hosted
  in-house within a controlled environment to ensure data privacy and compliance with infrastructure security requirements. As these internal tools and hosting
  infrastructure are proprietary and cannot be publicly released, the detailed step-by-step mathematical formalization and algorithmic descriptions provided in the
  preceding subsections are intended to ensure full reproducibility of the proposed approach using any compatible agentic AI development framework and large language model hosting
  environment. For the test cases author agent, the use of a language model with strong reasoning capabilities and a long context window is highly recommended; additionally, delegating component-wise test case generation to child sub-agents is advised to avoid context window saturation when processing platforms with racks comprising a large number of
  components.

\section{Experimental Setup}
\label{sec:experimental_setup}

This section describes the evaluation methodology used to assess the quality of the test plans generated by the proposed multi-agent framework. The evaluation employs a two-fold assessment strategy: an automated agents-as-judge protocol that leverages multiple independent large language models as reviewers, and a human expert annotation protocol that provides ground-truth quality assessments from domain engineers. Both protocols are applied to test plans generated for two production rack-scale datacenter platforms and repeated across multiple independent experimental replications to quantify evaluation stability.

\subsection{Evaluation Platforms and Test Plan Composition}

The framework is evaluated on two production AI datacenter platforms, hereafter referred to as Platform~A and Platform~B, to assess generalizability across distinct hardware architectures. Both platforms are rack-scale systems deployed in production AI infrastructure, comprising heterogeneous components spanning compute, power delivery, thermal management, networking, and system management subsystems. The platforms differ in component count, functional architecture, and self-healing validation document structure, providing complementary evaluation contexts.

For each platform, the multi-agent framework ingests the platform's self-healing validation document and bill of materials, and produces a structured test plan $\mathcal{P}$ as described in Section~\ref{methdology}. Each generated test plan contains two categories of test cases distinguished by provenance:

\begin{itemize}
    \item \textbf{Source-specification entries} ($\hat{\Phi}$): Test cases derived directly from the source self-healing validation document through the failure modes extractor agent, with gap-filled fields where applicable. These entries are expected to faithfully reproduce the original document content.
    \item \textbf{Agent-generated entries} ($\Phi_{\text{new}} \cup \mathcal{I}$): Test cases synthesized by the test cases author agent, including edge-case scenarios for under-covered components and cross-component interaction tests. These entries represent novel content produced by the framework.
\end{itemize}

Table~\ref{tab:platform_summary} summarizes the composition of the generated test plans for both platforms.

\begin{table}[!t]
\centering
\caption{Test plan composition for evaluation platforms}
\label{tab:platform_summary}
\small
\begin{tabular}{lcc}
\toprule
\textbf{Metric} & \textbf{Platform A} & \textbf{Platform B} \\
\midrule
Total test cases & 54 & 56 \\
Source-specification entries & 31 & 37 \\
Agent-generated entries & 23 & 19 \\
\bottomrule
\end{tabular}
\end{table}

\subsection{Agents-as-Judge Evaluation Protocol}
\label{sec:agents_as_judge}

The first evaluation dimension employs an automated multi-model review protocol in which three independent large language models serve as reviewers. This approach is motivated by two considerations: it provides a scalable, reproducible evaluation mechanism that can be applied consistently across platforms and replications, and the use of multiple reviewer models with distinct reasoning characteristics enables assessment of evaluation robustness through inter-model agreement analysis.

Three large language models from the Claude model family are selected as independent reviewers based on their demonstrated capabilities in structured reasoning, long-context comprehension, and domain-aware evaluation. Table~\ref{tab:reviewer_models} summarizes the key characteristics of each reviewer model. The selection spans two model generations (Claude~4.5 and Claude~4.6) and two capability tiers (Opus and Sonnet), providing architectural and behavioral diversity in reasoning patterns. Opus-class models employ extended chain-of-thought reasoning with higher computational budgets per token, making them well-suited for nuanced judgment tasks that require weighing multiple evaluation criteria simultaneously. The Sonnet-class model offers a complementary evaluation perspective through its optimization for balanced throughput and reasoning, which has been shown to produce distinct acceptance boundaries in structured evaluation tasks. All three models support context windows of at least 200K tokens, ensuring that the complete test plan, self-healing validation document, and evaluation instructions can be processed in a single context without truncation or retrieval augmentation. Each reviewer operates independently with no shared context, no access to other reviewers' assessments, and identical evaluation instructions, ensuring that inter-model agreement reflects genuine convergence in judgment rather than information leakage.

\begin{table}[!t]
\centering
\caption{Characteristics of large language models used as independent reviewers in the agents-as-judge evaluation protocol}
\label{tab:reviewer_models}
\small
\renewcommand{\arraystretch}{1.3}
\begin{tabular}{llccc}
\toprule
\textbf{Reviewer} & \textbf{Model} & \textbf{Family} & \textbf{Context} \\
\midrule
$\mathcal{R}_1$ & Opus 4.6 & Claude  & 200K  \\
$\mathcal{R}_2$ & Sonnet 4.6 & Claude  & 200K  \\
$\mathcal{R}_3$ & Opus 4.5 & Claude  & 200K \\
\bottomrule
\end{tabular}
\end{table}

The evaluation applies distinct criteria to source-specification entries and agent-generated entries, reflecting their different provenance and quality requirements. For source-specification entries $\hat{\phi} \in \hat{\Phi}$, each reviewer verifies extraction fidelity by comparing the \texttt{failure\_description} field against the original self-healing validation document. A source-specification entry is scored as:
\begin{equation}
  \text{score}_{\text{spec}}(\hat{\phi}) =
  \begin{cases}
  1 & \text{if } \hat{\phi}[\text{failure\_desc}] \text{ matches source} \\
  0 & \text{otherwise}
  \end{cases}
\end{equation}
For agent-generated entries $p \in \Phi_{\text{new}} \cup \mathcal{I}$, each reviewer independently evaluates whether the test case constitutes a valid, non-redundant hardware fault injection scenario. Each reviewer $\mathcal{R}_k$ produces a binary validity score $v_k(p) \in \{0, 1\}$ for $k \in \{1, 2, 3\}$, where $v_k(p) = 1$ indicates the reviewer judges the test case as valid and $v_k(p) = 0$ indicates rejection. Along with each score, the reviewer provides a textual rationale documenting the basis for acceptance or rejection, enabling post-hoc analysis of rejection patterns and systematic identification of disagreement axes across models.

The final validity determination for each agent-generated entry is computed through majority vote across the three reviewers:
\begin{equation}
\text{consensus}(p) =
\begin{cases}
\text{pass} & \text{if } \sum_{k=1}^{3} v_k(p) \geq 2 \\
\text{fail} & \text{otherwise}
\end{cases}
\label{eq:consensus}
\end{equation}
This consensus mechanism provides robustness against individual model biases while preserving sensitivity to genuine quality issues. A test case that two or more reviewers independently reject is classified as invalid, regardless of the third reviewer's assessment. This threshold was chosen to balance false positive risk (accepting invalid test cases) against false negative risk (rejecting valid test cases that a single reviewer mischaracterizes). The consensus mechanism additionally enables classification of rejection strength:
\begin{equation}
\text{rejection type}(p) =
\begin{cases}
\text{unanimous} & \text{if } \sum_{k=1}^{3} v_k(p) = 0 \\
\text{majority} & \text{if } \sum_{k=1}^{3} v_k(p) = 1
\end{cases}
\label{eq:rejection_type}
\end{equation}
where unanimous rejections indicate high-confidence invalid entries on which all reviewers converge, and majority rejections indicate contested entries where one reviewer accepted what the other two rejected.

To assess the stability of the automated evaluation, the agents-as-judge protocol is executed three times independently for each platform, yielding a total of six evaluation runs ($3 \text{ replications} \times 2 \text{ platforms}$). Each replication constitutes a complete, independent execution of the three-reviewer evaluation pipeline on the same generated test plan. Variation across replications captures the stochastic component of language model evaluation behavior, providing confidence intervals on reported metrics.

\subsection{Human Expert Evaluation Protocol}
\label{sec:human_eval}

The second evaluation dimension employs domain expert annotation to provide ground-truth quality assessments that serve as the definitive benchmark for test case validity. Unlike the automated protocol, human evaluation captures nuanced engineering judgment about the practical utility and executability of generated test cases within the specific hardware validation context.

\subsubsection{Annotator Selection}

Two human annotators are selected from systems integration engineers with direct experience in fault injection testing on datacenter platforms. Each annotator possesses domain expertise in the platform architecture, self-healing validation document conventions, and the validation management system into which generated test plans are imported. This expertise is essential for assessing whether generated test cases specify realistic injection methods, reference valid detection pathways, and describe actionable verification procedures. Both annotators independently review the complete generated test plan, enabling computation of inter-annotator agreement as a measure of evaluation reliability.
\subsubsection{Annotation Schema}

Each human annotator reviews every test case in the generated test plan and assigns one of two quality labels:

\begin{equation}
    \text{label}(p) \in \{\textsc{Accept}, \textsc{Reject}\}
\end{equation}

The two-level schema is defined as follows:
\begin{itemize}
    \item \textbf{\textsc{Accept}}: The test case is valid and can be imported directly into the validation management system without modification. The failure description, injection method, detection pathway, and verification procedure are all correct and complete.
    \item \textbf{\textsc{Reject}}: The test case is invalid and should not be included in the validation test plan. Rejection reasons include duplication of existing source-specification entries, reliance on application-level detection without hardware fault signals, post-repair verification rather than fault injection, or specification of physically untestable scenarios.
\end{itemize}

%% Evaluation Metrics subsection to be added after results analysis is complete.

\section{Results}
\label{sec:results}

\subsection{Agents-as-Judge}
\label{{sec:agents_as_judge}}
This section presents the quantitative results of the agents-as-judge evaluation protocol described in Section~\ref{sec:agents_as_judge}. The evaluation assesses the framework across multiple complementary dimensions: test plan coverage expansion, source-specification extraction fidelity, per-model and consensus validity rates, consensus agreement distribution, component-level validation outcomes, and inter-rater agreement among reviewer models.

\begin{figure}[!t]
    \centering
    \includegraphics[width=\columnwidth]{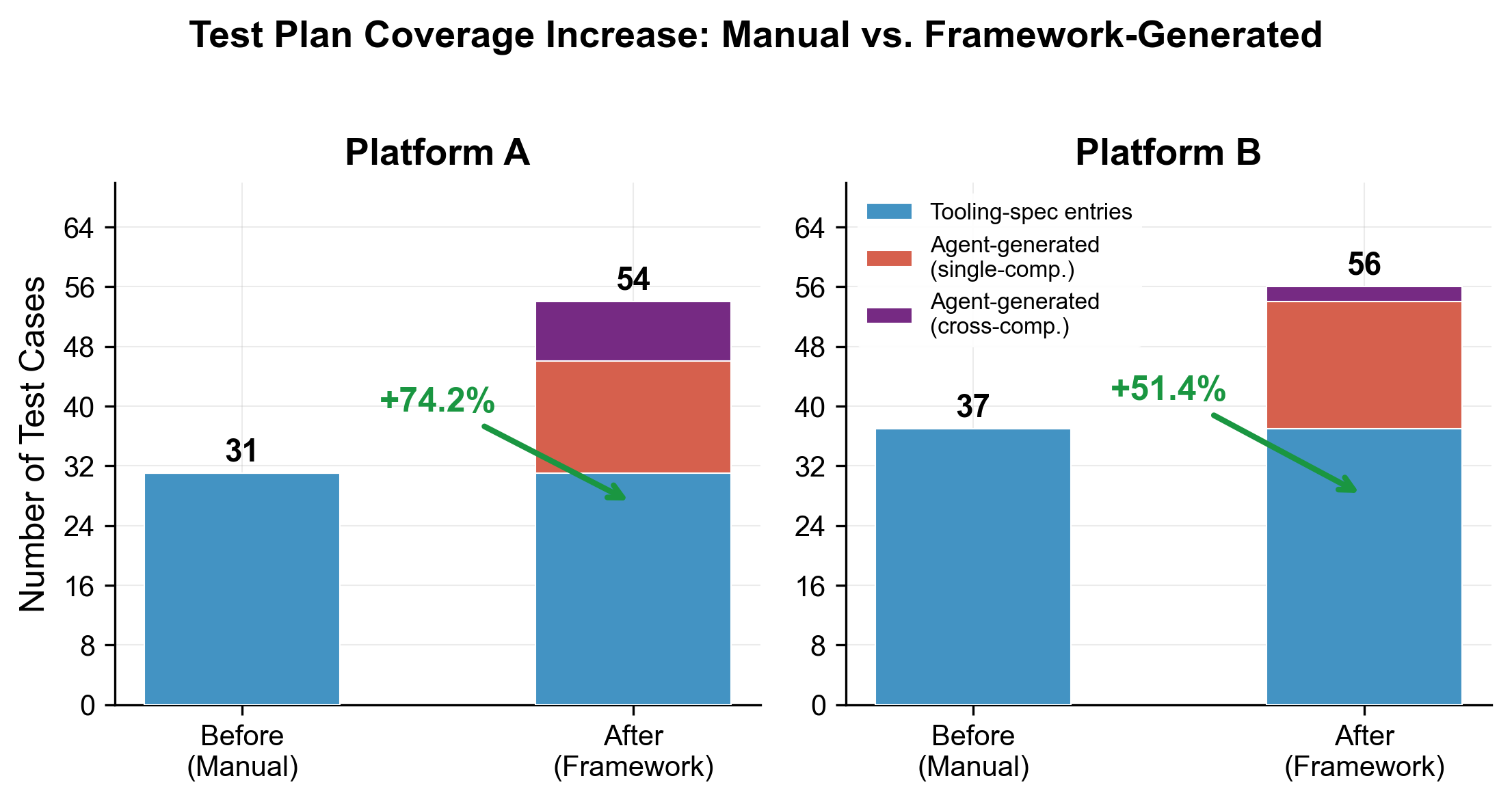}
    \caption{Test plan coverage increase from source-specification baseline to framework-generated output, decomposed by entry provenance.}
    \label{fig:coverage_increase}
\end{figure}

A primary objective of the framework is to expand test plan coverage beyond what manual specification-driven authoring provides. Table~\ref{tab:coverage_increase} and Figure~\ref{fig:coverage_increase} present the test plan size before and after applying the framework for both evaluation platforms. For Platform~A, the framework increased the test plan from 31 source-specification entries to 54 total test cases, a coverage expansion of $+74.2\%$. For Platform~B, the test plan grew from 37 to 56 entries, an increase of $+51.4\%$. The added test cases comprise both single-component edge cases for under-covered components and cross-component interaction tests that exercise fault propagation across functionally coupled subsystems.

\begin{table}[!t]
\centering
\caption{Test plan coverage increase: source-specification baseline versus framework-generated output}
\label{tab:coverage_increase}
\small
\renewcommand{\arraystretch}{1.3}
\begin{tabular}{lcc}
\toprule
\textbf{Metric} & \textbf{Platform A} & \textbf{Platform B} \\
\midrule
Before (source-specification only)    & 31 & 37 \\
After (framework-generated)   & 54 & 56 \\
Agent-generated (added)       & 23 & 19 \\
Coverage increase             & $+74.2\%$ & $+51.4\%$ \\
\bottomrule
\end{tabular}
\end{table}

\begin{figure}[!t]
    \centering
    \includegraphics[width=\columnwidth]{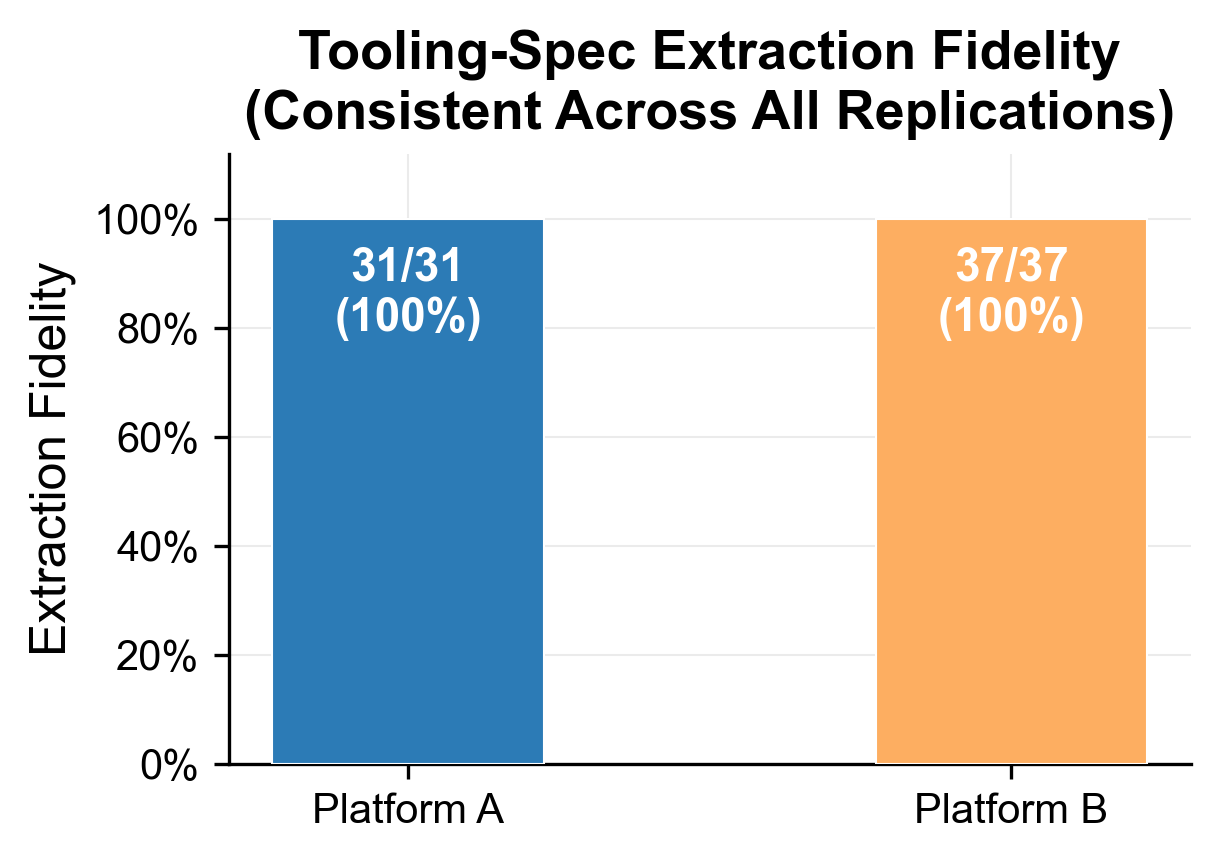}
    \caption{Source-specification extraction fidelity across evaluation platforms. All entries were correctly extracted in every replication.}
    \label{fig:extraction_fidelity}
\end{figure}

Before evaluating the quality of this novel agent-generated content, it is essential to verify that the framework faithfully preserves the information already present in the source specifications. Figure~\ref{fig:extraction_fidelity} reports the extraction fidelity for source-specification entries across both platforms. All three reviewer models unanimously confirmed that every source-specification entry was correctly extracted in all three replications: 31/31 for Platform~A and 37/37 for Platform~B, yielding 100\% extraction fidelity across all evaluation runs. This result establishes that the failure modes extractor agent introduces no information loss or distortion during the ingestion and normalization stage, providing a reliable foundation upon which the agent-generated entries build.

\begin{figure*}[!t]
    \centering
    \includegraphics[width=0.9\textwidth]{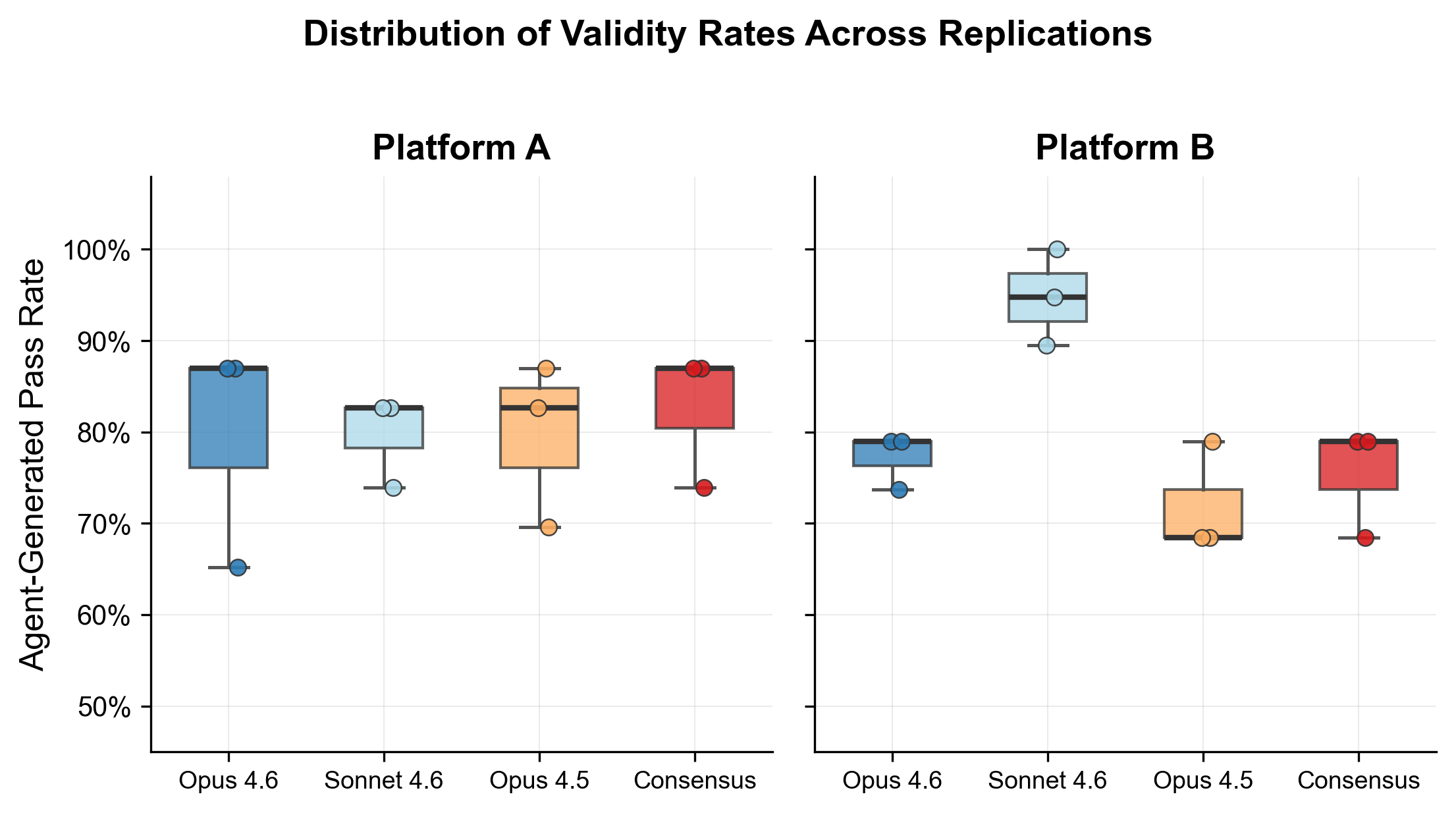}
    \caption{Distribution of agent-generated validity rates across three replications, per reviewer model and majority-vote consensus.}
    \label{fig:boxplot}
\end{figure*}

With extraction fidelity established, the evaluation turns to the quality of agent-generated test cases. Table~\ref{tab:per_model_pass_rates} summarizes the per-model and consensus pass rates across both platforms, reporting mean and standard deviation over three independent replications. Figure~\ref{fig:boxplot} visualizes the distribution of these rates, showing the median, interquartile range, and individual replication data points for each reviewer model and the consensus mechanism.

\begin{table*}[!t]
\centering
\caption{Per-model agent-generated pass rates across platforms and replications. Values report mean $\pm$ standard deviation over three independent replications.}
\label{tab:per_model_pass_rates}
\small
\renewcommand{\arraystretch}{1.3}
\begin{tabular}{lcccc}
\toprule
\textbf{Reviewer Model} & \multicolumn{2}{c}{\textbf{Platform A} ($n=23$)} & \multicolumn{2}{c}{\textbf{Platform B} ($n=19$)} \\
\cmidrule(lr){2-3} \cmidrule(lr){4-5}
 & \textbf{Mean $\pm$ $\sigma$ (\%)} & \textbf{Range (\%)} & \textbf{Mean $\pm$ $\sigma$ (\%)} & \textbf{Range (\%)} \\
\midrule
Claude Opus 4.6 ($\mathcal{R}_1$)   & $79.7 \pm 10.2$ & $65.2$--$87.0$  & $77.2 \pm 3.0$  & $73.7$--$78.9$ \\
Claude Sonnet 4.6 ($\mathcal{R}_2$) & $79.7 \pm 4.1$  & $73.9$--$82.6$  & $94.7 \pm 5.3$  & $89.5$--$100.0$ \\
Claude Opus 4.5 ($\mathcal{R}_3$)   & $79.7 \pm 7.4$  & $69.6$--$87.0$  & $71.9 \pm 6.1$  & $68.4$--$78.9$ \\
\addlinespace[2pt]
\midrule
Consensus ($\geq 2/3$)              & $82.6 \pm 6.1$  & $73.9$--$87.0$  & $75.4 \pm 6.1$  & $68.4$--$78.9$ \\
\bottomrule
\end{tabular}
\end{table*}

On Platform~A, all three reviewer models produced a mean pass rate of $79.7\%$, though with notably different variance profiles: Sonnet~4.6 exhibited the tightest spread ($\sigma = 4.1\%$), while Opus~4.6 showed the widest ($\sigma = 10.2\%$), reflecting differences in evaluation consistency across replications. The consensus mechanism, which requires agreement from at least two of three reviewers, achieved a mean pass rate of $82.6\% \pm 6.1\%$, exceeding each individual model's mean. This demonstrates the stabilizing effect of majority-vote aggregation: individual model variance is attenuated when a test case must convince multiple independent reviewers. On Platform~B, Sonnet~4.6 was the most permissive reviewer ($\mu = 94.7\%$), while Opus~4.5 was the most conservative ($\mu = 71.9\%$), producing a 22.8 percentage-point spread between the most and least lenient models. The consensus rate of $75.4\% \pm 6.1\%$ falls between these extremes, confirming that the majority-vote mechanism effectively mediates inter-model disagreement.

\begin{figure*}[!t]
    \centering
    \includegraphics[width=0.9\textwidth]{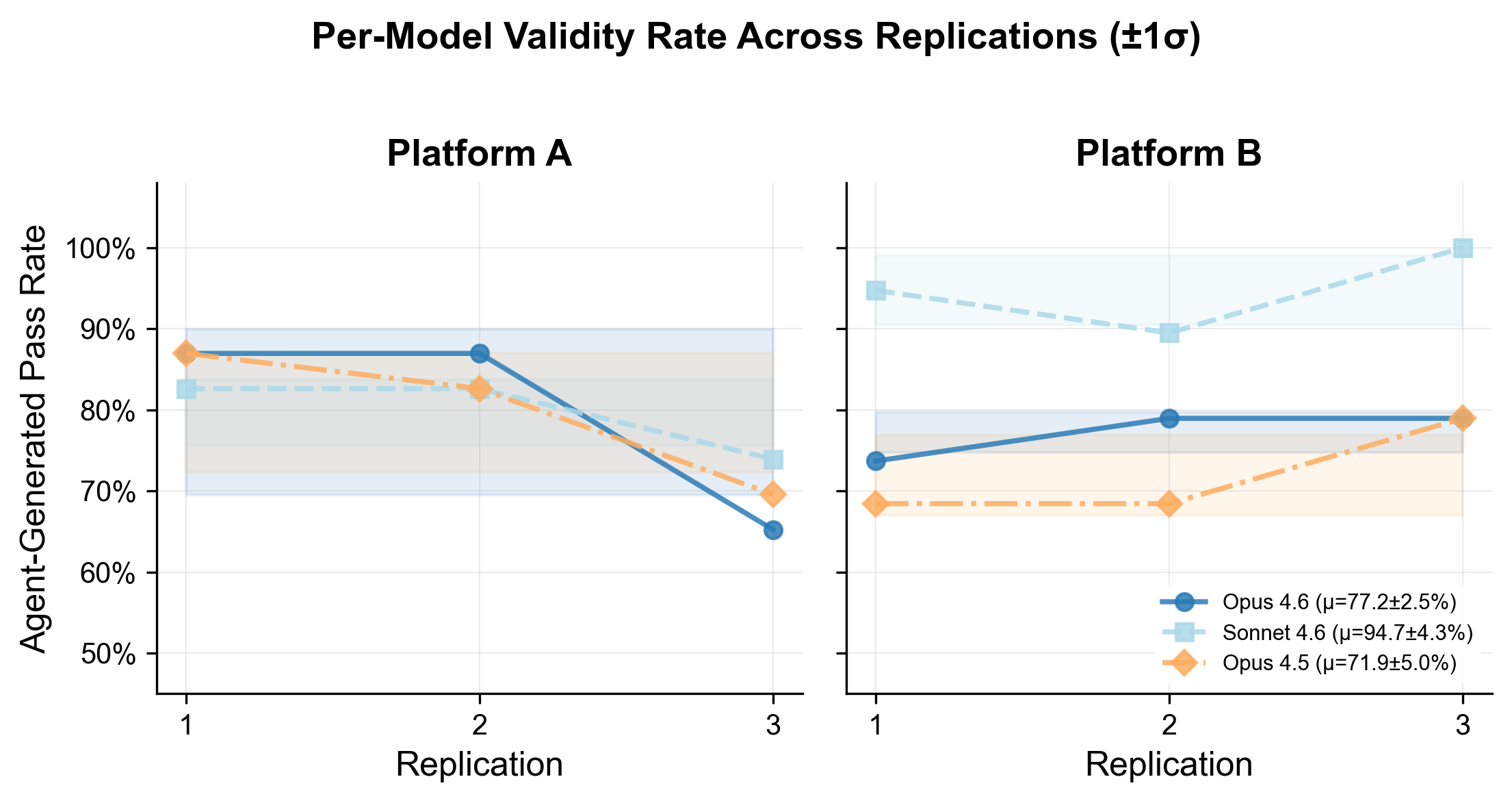}
    \caption{Per-model validity rate trajectories across replications with $\pm 1\sigma$ bands. Each model's band uses its own color to show individual evaluation stability.}
    \label{fig:model_variation}
\end{figure*}

Figure~\ref{fig:model_variation} presents the per-model pass rate trajectories across replications with per-model $\pm 1\sigma$ bands, providing a complementary view of evaluation stability. The visualization reveals that while individual replication outcomes vary, the models maintain consistent relative ordering across runs. The narrowness of the $\sigma$ bands for Sonnet~4.6 on both platforms suggests that this model applies the most deterministic evaluation criteria, whereas the wider bands for Opus~4.6 on Platform~A indicate greater sensitivity to stochastic variation in reasoning.

The overall consensus pass rate of approximately $75$--$83\%$ across both platforms indicates that roughly one in five agent-generated test cases is flagged for rejection by the automated review panel. This rejection rate is expected and, in fact, desirable within the intended human-in-the-loop workflow. The framework is designed to produce a comprehensive draft test plan that maximizes coverage recall surfacing failure scenarios that manual authoring would miss while accepting that a fraction of generated entries will require engineering judgment to filter. The alternative, a conservative generation strategy that only produces entries all models would accept, would sacrifice the novel edge-case scenarios that constitute the framework's primary value. The rejected entries predominantly fall into identifiable categories, post-repair verification procedures, connectivity audits, and physically untestable scenarios providing clear, actionable feedback that domain engineers can process efficiently during review. This positions the framework not as a replacement for human expertise, but as a force multiplier: it reduces the authoring burden from days of manual specification review to hours of draft refinement, while surfacing failure modes and component interactions that manual processes systematically overlook.

\begin{figure*}[!t]
    \centering
    \includegraphics[width=0.9\textwidth]{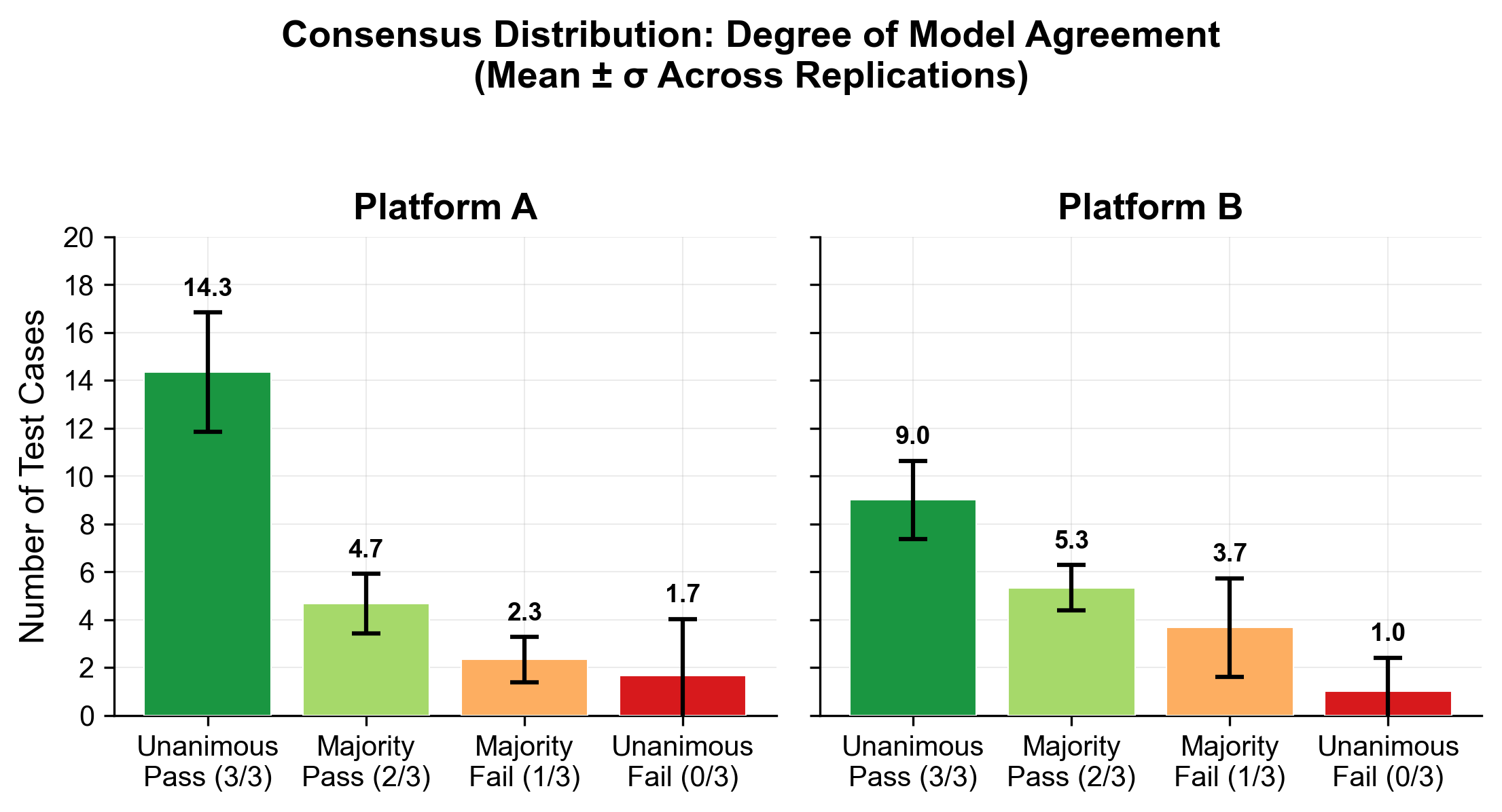}
    \caption{Consensus distribution showing degree of model agreement on agent-generated entries. Values report mean $\pm \sigma$ across three replications.}
    \label{fig:consensus_dist}
\end{figure*}

To understand the strength of reviewer agreement behind these rates, Figure~\ref{fig:consensus_dist} decomposes the consensus outcomes by degree of model agreement. On Platform~A, a mean of 14.3 out of 23 agent-generated entries received unanimous acceptance across replications, indicating that the majority of generated test cases are unambiguously valid. Only 1.7 entries on average received unanimous rejection, corresponding to clearly identifiable categories such as post-repair verification procedures that do not constitute fault injection tests. On Platform~B, 9.0 entries received unanimous acceptance out of 19, with only 1.0 receiving unanimous rejection on average. The contested region entries receiving majority pass or majority fail represents cases where at least one reviewer model disagreed with the others, reflecting genuine ambiguity in test case validity rather than clear-cut quality issues. The small size of the unanimous fail category on both platforms ($1.7$ and $1.0$ respectively) indicates that the framework rarely produces entries that all reviewers agree are invalid, reinforcing that most generated content represents reasonable validation scenarios even when individual models disagree on boundary cases.

\begin{figure*}[!t]
    \centering
    \includegraphics[width=0.9\textwidth]{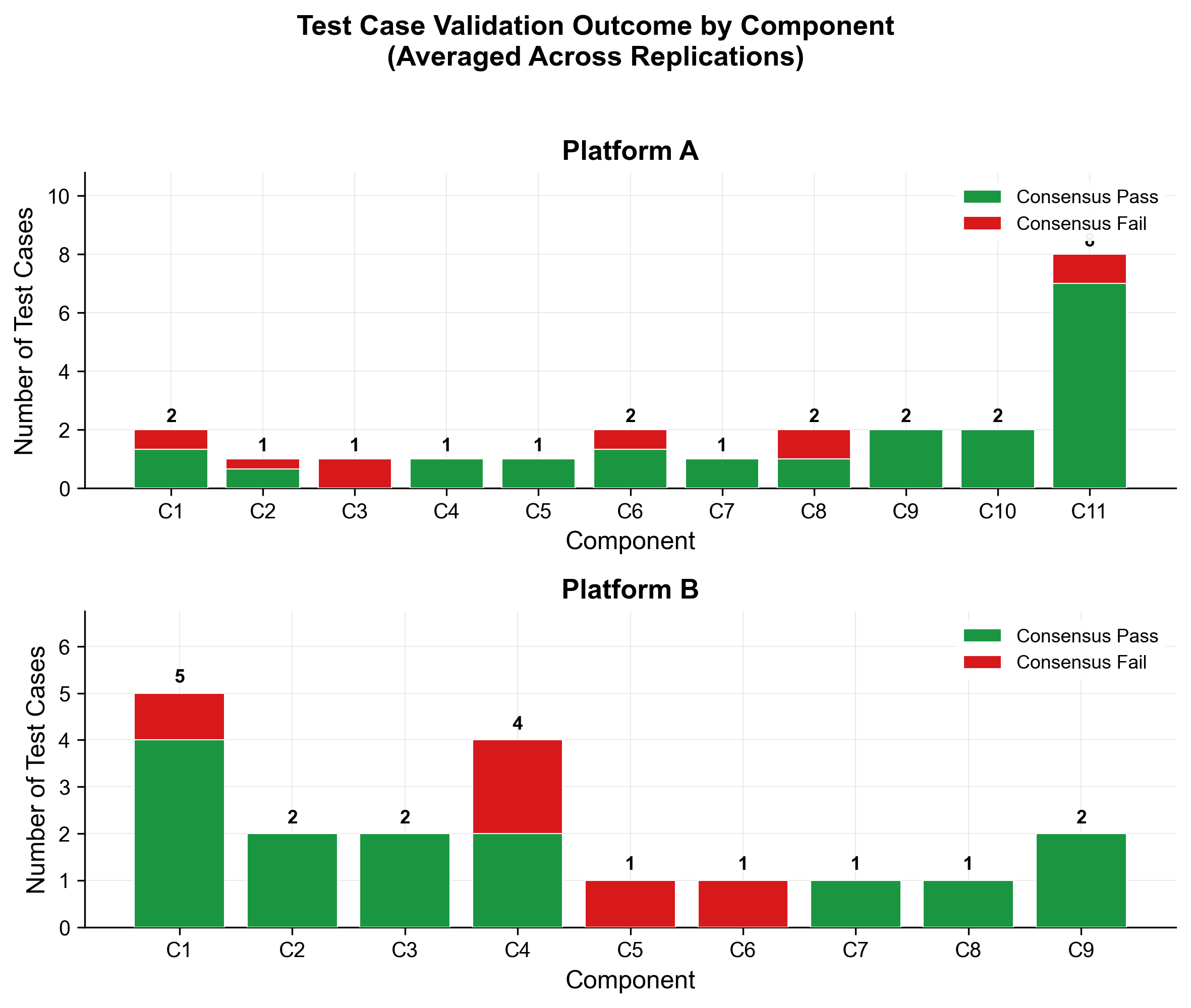}
    \caption{Test case validation outcome by component (consensus pass vs. fail), averaged across three replications.}
    \label{fig:stacked_pass_fail}
\end{figure*}

Breaking down validation outcomes at the component level provides further insight into where the framework performs strongest and where targeted refinement would be most impactful. Figure~\ref{fig:stacked_pass_fail} presents the consensus pass and fail counts decomposed by component, averaged across three replications. On Platform~A, 8 of 11 component categories achieved consensus pass rates above $80\%$, with several reaching $100\%$. On Platform~B, 6 of 9 component categories achieved $100\%$ pass rates, with failures concentrated in components where the framework generated test cases involving physical-only injection methods or component-specific addressing schemes that reviewers deemed insufficiently automatable. This concentration of rejections in a small number of components suggests that targeted prompt refinement for these specific domains would yield disproportionate quality improvements, rather than indicating a systemic quality issue across the framework's output.

\begin{figure}[!t]
    \centering
    \includegraphics[width=\columnwidth]{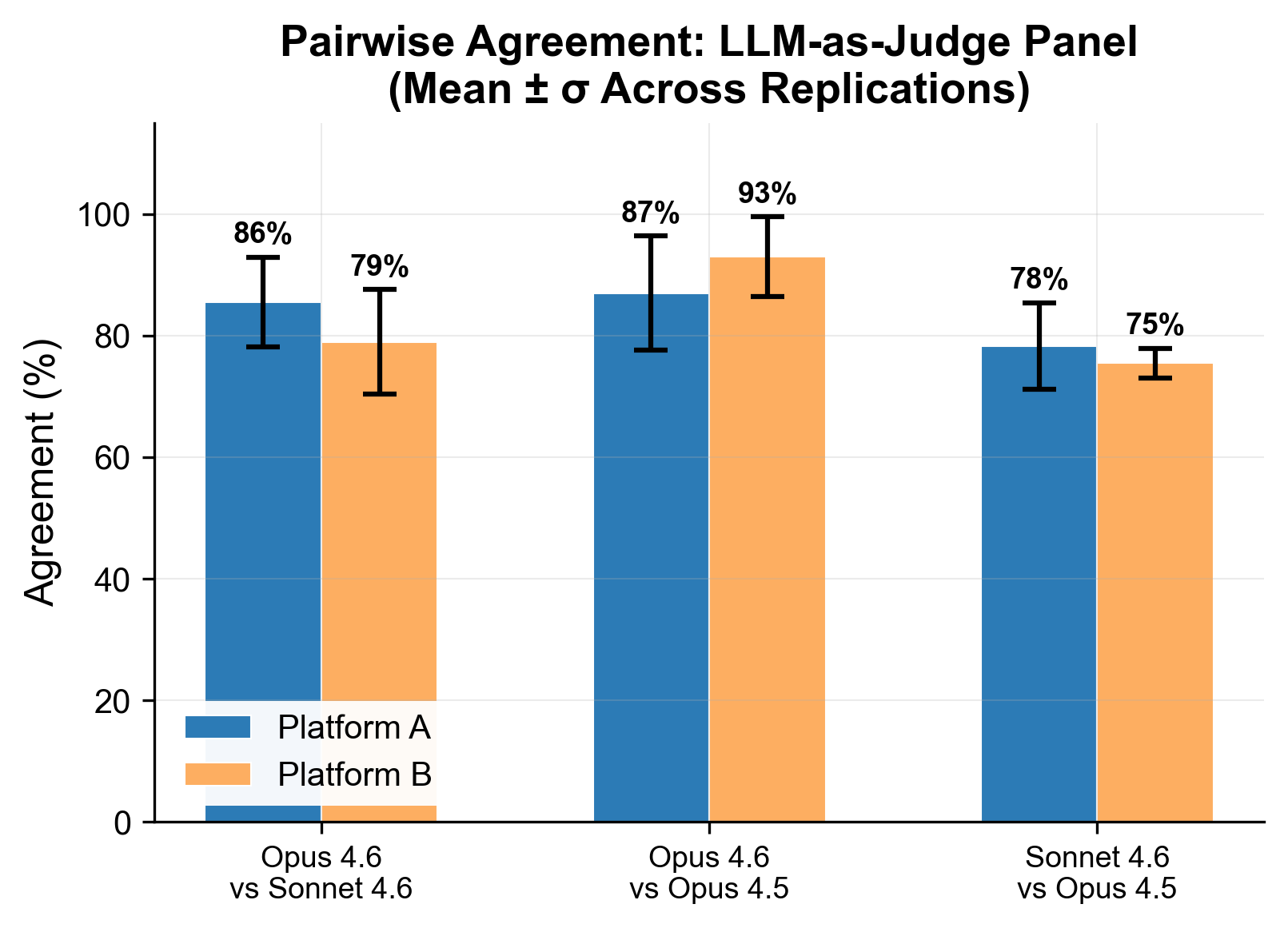}
    \caption{Pairwise agreement rates between reviewer models, reported as mean $\pm \sigma$ across three replications.}
    \label{fig:pairwise_agreement}
\end{figure}

Finally, to assess the reliability of the multi-model evaluation panel itself, Figure~\ref{fig:pairwise_agreement} reports pairwise agreement rates between all reviewer model pairs across both platforms. Agreement rates range from $75\%$ to $93\%$ across all pairs and platforms, with mean values consistently above $78\%$. The highest agreement on Platform~B was observed between Opus~4.6 and Opus~4.5 ($93\%$), which share the same model tier (Frontier) and exhibit similar strictness patterns on configuration audit and maintenance check entries. On Platform~A, all three pairs achieved agreement above $85\%$, indicating strong convergence in evaluation behavior. These agreement levels are consistent with the range reported in the literature for human expert inter-annotator agreement on engineering judgment tasks, where perfect agreement is neither expected nor desirable, some degree of disagreement reflects legitimate differences in interpreting boundary cases. The consistently high pairwise agreement supports the reliability of the agents-as-judge protocol, while the moderate disagreement on specific entry types highlights precisely the ambiguous categories where human expert judgment is most valuable further reinforcing the human-in-the-loop design philosophy in which the automated panel identifies high-confidence valid and invalid entries with strong agreement, while flagging contested entries for expert review.

\subsection{Human Expert Evaluation}

This section presents the results of the human expert evaluation protocol described in Section \ref{sec:human_eval}. While the agents-as-judge protocol provides statistically stable, reproducible assessment through multiple replications, human evaluation surfaces practical engineering constraints that automated reviewers cannot assess, including physical safety, testability feasibility, and organizational coverage decisions. Together, the two evaluation dimensions provide complementary validation of the framework's output quality.

Table~\ref{tab:human_eval} summarizes the human expert evaluation outcomes for agent-generated test cases on both platforms. Each platform was reviewed by a domain engineer with direct experience in fault injection testing on that specific hardware architecture.

\begin{table}[!t]
    \centering
    \caption{Human expert evaluation of agent-generated test cases}
    \label{tab:human_eval}
    \small
    \renewcommand{\arraystretch}{1.3}
    \begin{tabular}{lcc}
        \toprule
        \textbf{Metric} & \textbf{Platform A} ($n = 23$) & \textbf{Platform B} ($n = 19$) \\
        \midrule
        Accept & 18 (78.3\%) & 13 (68.4\%) \\
        Reject & 5 (21.7\%) & 6 (31.6\%) \\
        \bottomrule
    \end{tabular}
\end{table}

\begin{figure}[!t]
    \centering
    \includegraphics[width=\columnwidth]{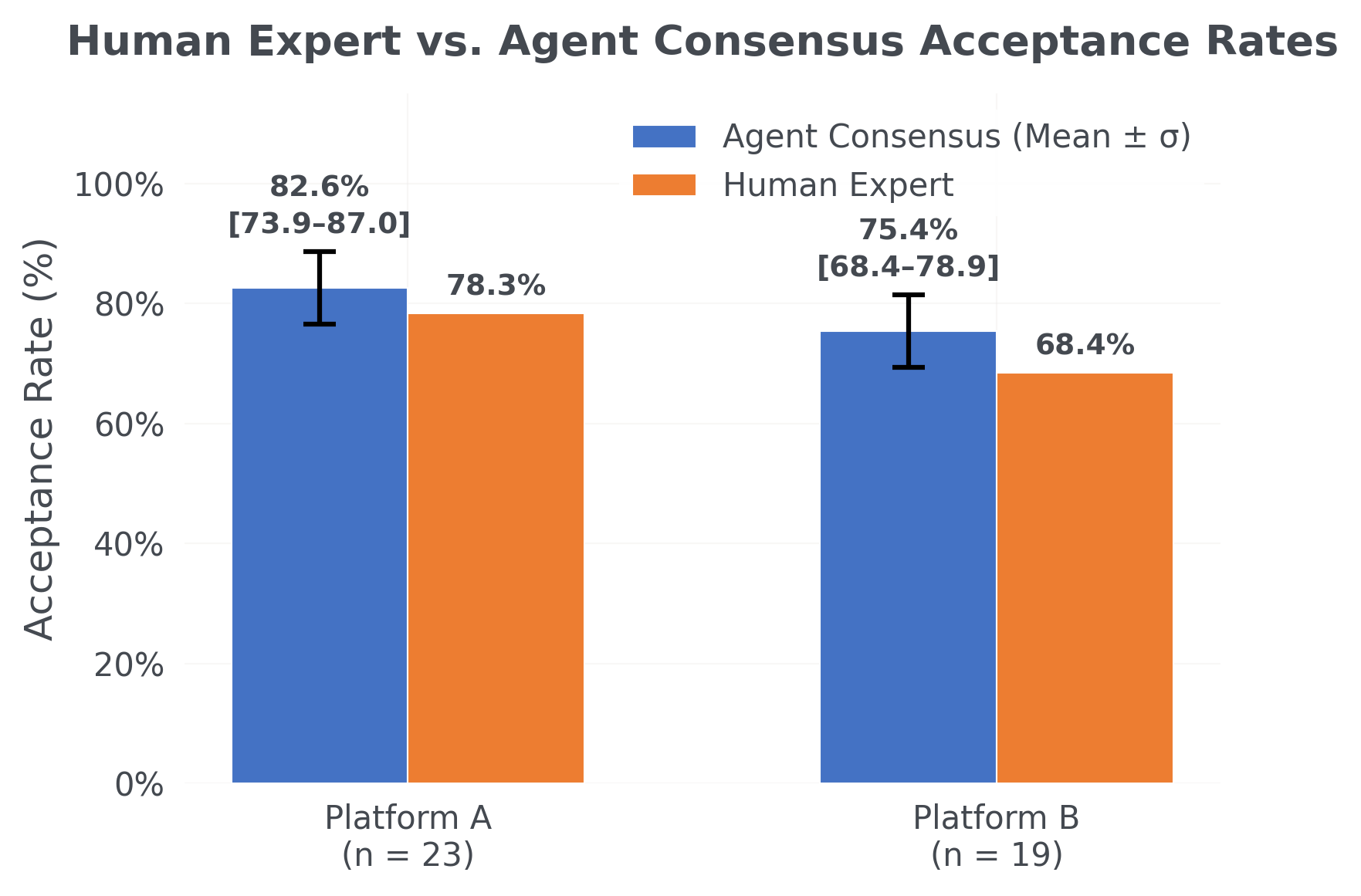}
    \caption{Human expert vs.\ agent consensus acceptance rates. Human rates fall within agent replication ranges on both platforms, confirming evaluation calibration. Error bars on agent bars represent $\pm 1\sigma$ across three replications.}
    \label{fig:human_vs_agent}
\end{figure}

Fig.~\ref{fig:human_vs_agent} compares human acceptance rates against the agent consensus results from Section \ref{{sec:agents_as_judge}}. Platform~A's human acceptance rate of 78.3\% falls within the agent consensus replication range of 73.9--87.0\%, and Platform~B's rate of 68.4\% matches the lower bound of the agent range of 68.4--78.9\%. This alignment provides strong evidence that the agents-as-judge protocol is well-calibrated against domain expert judgment. Human experts are slightly stricter overall. Their rates sit at or below the agent consensus mean (82.6\% and 75.4\% respectively), which reflects the expected direction: domain engineers apply practical constraints such as physical safety and test redundancy that language model reviewers cannot fully assess.

Cross-component interaction tests were validated strongly by human experts: Platform~A achieved 8/8 (100\%) human acceptance. Platform~B's single cross-component rejection was due to a physical safety concern (risk of partially disconnecting a power whip under load), not a quality or reasoning deficiency.

The rejection taxonomy reveals that all 11 rejected entries across both platforms fall into four well-defined, actionable categories. None represent hallucinations or fundamental reasoning failures. Table~\ref{tab:rejection_taxonomy} summarizes these categories with their engineering dispositions.

\begin{table*}[!t]
    \centering
    \caption{Rejection taxonomy and engineering disposition. All rejections are scope decisions, not quality failures.}
    \label{tab:rejection_taxonomy}
    \small
    \renewcommand{\arraystretch}{1.3}
    \begin{tabular}{p{3.8cm}ccp{4.2cm}p{4.5cm}}
        \toprule
        \textbf{Rejection Class} & \textbf{Plat.~A} & \textbf{Plat.~B} & \textbf{Engineering Verdict} & \textbf{Implication} \\
        \midrule
        Redundancy with existing test coverage & 5 & 0 & Merge into parent test case & Framework independently identified failure domains already in spec \\
        Detection path covered by broader monitor & 0 & 3 & No action: confirms existing monitoring sufficiency & Coverage validation, not generation error \\
        Physical safety risk during injection & 0 & 2 & Defer to controlled maintenance window or simulation & Constraint only human experts can assess \\
        Hardware physically untestable & 0 & 1 & Remove: overestimated testability & Actionable feedback for prompt refinement \\
        \midrule
        \textbf{Total} & \textbf{5} & \textbf{6} & & \\
        \bottomrule
    \end{tabular}
\end{table*}

Three findings emerge from this analysis. First, zero rejected entries contain fabricated components or nonsensical failure modes; every rejection describes a real hardware component and a plausible fault scenario. The engineer's judgment is whether to \emph{include} the test case in the automated suite, not whether it is \emph{technically valid}. Second, the rejection categories are precisely those that require human-in-the-loop judgment: physical safety assessment, organizational scope decisions, and redundancy with existing coverage that requires institutional knowledge to identify. Automated reviewers flagged the same categories as humans (post-repair procedures, physical-only injection) but cannot assess the safety dimension, empirically validating the human-in-the-loop architecture. Third, Platform~A's five redundancy rejections demonstrate that the framework independently derived failure scenarios already present in the self-healing validation document, a form of coverage self-validation that confirms the soundness of the system's domain reasoning even in cases where the output is ultimately filtered.

\section{Conclusion}
\label{conclusion}
This paper presented a generative AI multi-agent framework designed to automate the authoring of hardware validation test plans for large-scale AI datacenter platforms. By decomposing the workflow into specialized agents for domain classification, failure mode extraction, and test case synthesis, the proposed architecture successfully bridged the gap between heterogeneous engineering inputs, specifically BOMs, self-healing validation documents, and structured, actionable validation procedures. Evaluation on two production rack-scale AI platforms demonstrated significant coverage expansions of 74.2\% and 51.4\% over manual baselines while maintaining 100\% extraction fidelity for existing specifications.

Despite these advancements, the study highlights several limitations inherent to the current multi-agent approach. While the generative agents successfully synthesize technically plausible fault scenarios, human expert evaluation revealed that the system occasionally proposes tests involving hardware that is practically untestable in a deployed environment. Furthermore, the framework currently lacks deep contextual awareness of organizational testing scopes, sometimes generating redundant scenarios or post-repair verification procedures rather than genuine fault injections. These boundary cases underscore the continued necessity of human-in-the-loop oversight to filter outputs based on physical constraints and operational realities that language models cannot fully assess yet.

Future work will focus on refining the framework to address these practical constraints and broaden its applicability. A primary next step involves integrating explicit physical safety guardrails and testability heuristics into the generation prompts to minimize non-actionable outputs and reduce the human review burden. Additionally, extending the architecture to ingest firmware specifications and system-level telemetry could enable closed-loop validation, where test execution results dynamically feed back into the agents to continuously refine generation accuracy. Exploring the application of this methodology to other complex cyber-physical domains will further validate the scalability of multi-agent architectures in automating rigorous engineering compliance and hardware validation pipelines.


\newpage
\bibliographystyle{IEEEtran}
\bibliography{references}

@article{yu2026survey,
  title={A survey on failure analysis and fault injection in AI systems},
  author={Yu, Guangba and Tan, Gou and Huang, Haojia and Zhang, Zhenyu and Chen, Pengfei and Natella, Roberto and Zheng, Zibin and Lyu, Michael R},
  journal={ACM Transactions on Software Engineering and Methodology},
  volume={35},
  number={1},
  pages={1--42},
  year={2026},
  publisher={ACM New York, NY}
}

@article{cinque2025cosmos,
  title={COSMOS: A Fault Injection Framework to Assess Hardware-Assisted Hypervisors},
  author={Cinque, Marcello and Cotroneo, Domenico and De Rosa, Giuseppe and De Simone, Luigi and Farina, Giorgio},
  journal={IEEE Transactions on Dependable and Secure Computing},
  year={2025},
  publisher={IEEE}
}

@article{opara2025chaos,
  title={Chaos Engineering 2.0: A Review of AI-Driven, Policy-Guided Resilience for Multi-Cloud Systems},
  author={Opara, Lasbrey Chibuzo and Akatakpo, Ogheneruemu Nathaniel and Ironuru, Ifeanyi Charles and Anyaene, Kingsley and Enobakhare, Benjamin Osaze},
  journal={Journal of Computer, Software, and Program},
  volume={2},
  number={2},
  pages={10--24},
  year={2025}
}

@article{yu2025systematic,
  title={A Systematic Literature Review on Fault Injection Testing of Microservice Systems},
  author={Yu, Senyao and Wu, Huayao and Niu, Xintao and Nie, Changhai},
  journal={IEEE Transactions on Services Computing},
  year={2025},
  publisher={IEEE}
}

@inproceedings{pourreza2023survey,
  title={A survey of faults and fault-injection techniques in edge computing systems},
  author={Pourreza, Maryam and Narasimhan, Priya},
  booktitle={2023 IEEE International Conference on Edge Computing and Communications (EDGE)},
  pages={63--71},
  year={2023},
  organization={IEEE}
}

@article{Ray2026TRiSM,
  title={A Review of TRiSM Frameworks in Artificial Intelligence Systems: Fundamentals, Taxonomy, Use Cases, Key Challenges and Future Directions},
  author={Ray, Partha Pratim},
  journal={Expert Systems},
  volume={43},
  number={3},
  pages={e70213},
  year={2026},
  publisher={Wiley Online Library}
}

@article{zhang2026Infrastructure,
  title={Higher-Order Network Approach for Modeling Cascading Failure in Urban Critical Infrastructure},
  author={Zhang, Li and Yang, Xingang and Pan, Aiqiang and Yang, Liu and Zhang, Yufan and Liu, Chang and Wu, Jiansong},
  journal={ASCE-ASME Journal of Risk and Uncertainty in Engineering Systems, Part A: Civil Engineering},
  volume={12},
  number={2},
  pages={04026002},
  year={2026},
  publisher={American Society of Civil Engineers}
}

@article{morgan2025digital,
  title={Digital Twin-Driven Cybersecurity for 5G/6G-Enabled Electric Vehicle Charging Infrastructure: A Review},
  author={Morgan, Ernest Fiko and Ali, Mohd Hasan},
  journal={Energies},
  volume={18},
  number={22},
  pages={6048},
  year={2025},
  publisher={MDPI}
}

@article{Hao2026Survey,
  title={DNN partitioning for cooperative inference in edge intelligence: Modeling, solutions, toolchains},
  author={Hao, Yuntao and Ding, Nan and Xia, Weiguo and Ge, Hongwei and Xu, Li},
  journal={ACM Computing Surveys},
  volume={58},
  number={8},
  pages={1--34},
  year={2026},
  publisher={ACM New York, NY}
}

@article{Uslu2026Orchestration,
  title={Next-Generation Management and Orchestration for 6G: Emerging Trends, Architectures, and Challenges},
  author={Uslu, Ege Erberk and Ozturk, Bunyamin and Gorkemli, Burak and Soyturk, Mesut and Cihan, Salih Bilge and Gunduzhev, Aycan Ramazan and Dagdeviren, Orhan},
  journal={IEEE Open Journal of the Communications Society},
  year={2026},
  publisher={IEEE}
}

@article{ivanov2026agentic,
  title={Agentic digital twins: bridging model-based and AI-driven decision-making support for a new era of supply chain and operations management},
  author={Ivanov, Dmitry},
  journal={International Journal of Production Research},
  pages={1--17},
  year={2026},
  publisher={Taylor \& Francis}
}

@article{Liu2026STAR,
  title={STAR: Steelmaking Task-Aware Routing for Multi-Agent LLM Expert Systems},
  author={Liu, Wenyuan and Huang, Chengyan and Wang, Songlei and Wang, Lin and Meng, Fanjie and Li, Minghui and Zhang, Haoning and Zheng, Qiang},
  journal={Electronics},
  volume={15},
  number={4},
  pages={720},
  year={2026},
  publisher={MDPI}
}

\end{document}